\documentstyle[12pt,epsf,here]{article}

\begin{document}

\newcommand{\Figurebb}[9]{
\begin{figure}[H]\begin{center}
\leavevmode
\epsfysize=#7cm
\epsfbox[#2 #3 #4 #5]{#6}
\par
\parbox{#8cm}{
\caption[figure]{\renewcommand{\baselinestretch}{0.8} \small
                                           \hspace{-0.3truecm}#9}
\label{#1}}
\end{center}
\end{figure}
}
\def\fig#1{Fig.~\protect\ref{#1}}
\def\hom{\hbar\omega}
\def\bs{\bigskip}
\def\ms{\medskip}
\def\siml{\,\hbox{\kern.1em \lower.6ex \hbox{$\sim$} \kern-1.12em
          \raise.6ex \hbox{$<$} }}
\def\simg{\,\hbox{\kern.1em \lower.6ex \hbox{$\sim$} \kern-1.12em
          \raise.6ex \hbox{$>$} }}
\def\eq#1{(\protect\ref{#1})}
\def\eqq#1#2{(\ref{#1},\ref{#2})}
\def\be{\begin{equation}}
\def\ee{\end{equation}}
\def\bea{\begin{eqnarray}}
\def\eea{\end{eqnarray}}
\def\M{{\cal M}}

\hfill TPR-98-18

\vspace{2.0cm}

\centerline{\bf \LARGE Uniform trace formulae for SU(2) and SO(3)}
\vspace{0.5cm}
\centerline{\bf \LARGE symmetry breaking}
\vspace{2.0cm}
\centerline{\large M.\ Brack, P.\ Meier and K.\ Tanaka$^*)$}
\vspace{0.4cm}

\noindent Institut f\"ur Theoretische Physik, Universit\"at
          Regensburg, D-93040 Regensburg, Germany

\vspace{2.4cm}
\centerline{\bf Abstract}
\vspace{0.5cm}

We develop uniform approximations for the trace formula for non-integrable
systems in which SU(2) symmetry is broken by a non-linear term of the
Hamiltonian. As specific examples, we investigate H\'enon-Heiles type
potentials. Our formalism can also be applied to the breaking of SO(3) symmetry
in a three-dimensional cavity with axially-symmetric quadrupole deformation.

\vspace{0.5cm}


\vspace{1.5cm}

\centerline{November 5, 1998}

\vspace{2.0cm}

\centerline{J. Phys. {\bf A}, in print}

\vfill

$^*)$ {\it Present address:} Department of Physics, University of Alberta,
      Edmonton, Alberta, Canada T6G 2J1.

\newpage

\section{Introduction}

Systems with mixed classical dynamics have so far offered the most difficult
problems in the attempts of a semiclassical quantization in terms of periodic
orbits \cite{gutz,bb,bt,book}. These problems have mainly three origins: 1) the
existence of continuous symmetries that make (some of) the periodic orbits
non-isolated, 2) bifurcations of stable orbits, and 3) the proximity of a
higher symmetry that is reached by letting a continuous parameter go to zero.
In all three cases, the original trace formula of Gutzwiller \cite{gutz} cannot
be used because the stationary-phase integration transverse to the periodic
orbits, that is used in its derivation, is not justified and leads to
divergences. By now, the problems connected to 1) and 2) are essentially
solved. Besides fully integrable systems \cite{bb,bt}, non-integrable systems
with various kinds of continuous symmetries can also be treated by properly
extended versions of the Gutzwiller theory \cite{stma,gsym}. Considerable
progress has also been made recently in the treatment of bifurcations, after
earlier indications how to go beyond the simplest saddle-point integration
\cite{bb,bt,rich}. Sieber and Schomerus \cite{bif,siel} have systematically
developed uniform approximations to the most common types of bifurcations,
expanding the action integrals in the neighborhood of a bifurcation point into
normal forms in phase space \cite{ozha}. The resulting trace formulae
interpolate continuously between the appropriate Gutzwiller limits that are
sufficiently far away from the bifurcation points, where the stationary-phase
integration can be applied.

In the present paper, we shall be concerned with the third type of problems
which arise from the breaking of a given symmetry through a continuous
parameter in the Hamiltonian. Let us start from an integrable system described
by a Hamiltonian $H_0$ which possesses a certain continuous symmetry. As a
consequence of this symmetry, the periodic orbits in the classical system occur
in degenerate families living on $N$-tori in phase space, where $N$ is the
number of degrees of freedom of the integrable system. We now perturb the
system by adding to it a term that breaks the symmetry:
\begin{equation}\label{Hpert}
 H = H_0 + \epsilon H_1\,.
\end{equation}
Here $\epsilon$ is a continuous dimensionless parameter which in the following
we may also call a ``deformation''. Due to the symmetry breaking, some (or all)
of the rational tori containing the periodic orbit families are broken up into
orbits that have a lesser degree of degeneracy than those of $H_0$, or are
completely isolated. The system \eq{Hpert} will in general exhibit mixed
classical dynamics, and if $H_1$ breaks all continuous symmetries of $H_0$, it
will become chaotic for large values of $\epsilon$ and for large energies,
where the standard Gutzwiller trace formula can be applied. However, for small
$\epsilon$ the amplitudes in this formula become very large; they actually
diverge for $\epsilon\rightarrow 0$. This is due to the fact that although the
rational tori are broken up for $\epsilon>0$, the periodic orbits are still not
sufficiently isolated as long as their perturbed actions differ by less than
$\hbar$, and consequently the stationary-phase integration transverse to the
orbits fails as mentioned above. One then has to find more accurate ways of
performing the trace integration over the semiclassical Green's function; in
principle, closed non-periodic orbits will thereby also contribute
significantly to the result \cite{mon}.

In the limit of small perturbations, $\epsilon \ll 1$, classical perturbation
theory may be used to derive trace formulae with finite amplitudes that yield
the correct limit for $\epsilon\rightarrow 0$. Generalizing earlier attempts
\cite{OdA,URJ}, Creagh \cite{crp} has recently developed a scheme to derive
perturbative trace formulae for the breaking of arbitrary continuous
symmetries, including e.g.\ SO(3) (spherically symmetric potentials) or SU($N$)
(harmonic oscillators in $N$ dimensions). Applications of this approach have
been presented in Refs.\ \cite{mbc,clus,kaor,bcl}. The results successfully
describe the transitions from higher to lower (or no) symmetry for small or
moderate deformations $\epsilon$, but they eventually fail when the
perturbative regime is exceeded. In the limit of large perturbation, $\epsilon
\gg 1$, one would like to recover the Gutzwiller trace formula \cite{gutz} for
isolated orbits, or its corresponding extension \cite{stma,gsym} if some
continuous symmetries are left. A closed form of an approximation that yields
this limit as well as the correct trace formula of the integrable system $H_0$
for $\epsilon\rightarrow 0$ is called a uniform approximation, in analogy to
the uniform approximations that interpolate continuously across bifurcations.

Tomsovic, Grinberg, and Ullmo \cite{TGU} have recently derived
a uniform approximation for the breaking of U(1) symmetry in a two-dimensional
system. Their result is quite general and applies to all systems where the
rational tori are broken into pairs of stable and unstable isolated orbits.
No analogous result is known to us for the breaking of a higher symmetry in any
dimension. We will discuss the approach of Ref.\ \cite{TGU} briefly in Sect.\
II and rederive it from the perturbative limit in a heuristic way that is
suitable for an extension to higher symmetries of $H_0$.

In this paper, we derive uniform approximations to perturbed harmonic
oscillators in two dimensions, where $H_0$ has SU(2) symmetry (Sect.\ III). We
furthermore apply one of the results to a three-dimensional cavity with small
axially-symmetric quadrupole deformations, where one starts from SO(3) symmetry
(Sect.\ IV).

Our aim is not the full quantization of these systems, but the description of
their gross-shell properties which are determined by the shortest orbits
\cite{book,stma}. The use of the periodic orbit theory to describe shell effects
in many-fermion systems in terms of a few short orbits has found nice
applications, e.g., in nuclei, for ground-state deformations \cite{nuc}
and the mass asymmetry of fission \cite{fis}; in metal clusters, for
supershells \cite{nis} and (using the perturbative approach \cite{crp}) their
modifications due to deformations \cite{mbc,clus} and magnetic fields
\cite{kaor}; and in semiconductor quantum dots, for magnetization 
\cite{URJ,kao2} and conductance fluctuations \cite{clus,qdot}.

\section{Recapitulation of U(1) breaking}

We start by recapitulating the work of Tomsovic {\it et al.}\ \cite{TGU}
for the breaking of U(1) symmetry in a two-dimensional system. We shall
re-derive here their result in the simplified version given by Sieber
\cite{siel}, using a heuristic way which will be generalized in the later
sections to systems with higher symmetries.

We restrict ourselves to the most frequently occurring case that a periodic
orbit family on a 2-torus is broken into a non-degenerate pair of stable
and unstable isolated orbits. In the two-dimensional trace integral, one
integration is performed exactly along the orbits as usual. The space variable
$q$ transverse to the orbit can always be mapped onto a variable $\phi$ which
is cyclic in $[0,2\pi)$ and as a function of which the action shift is
proportional to $\cos(\phi)$ (``pendulum mapping'' in \cite{TGU}). Hence the
contribution of an orbit family to the trace formula is
\begin{equation}\label{U1int}
  \delta g = \Re e \left\{ e^{i\Phi_0} \frac{1}{2\pi} \int_0^{2\pi}A(\phi)
             {\cal J}(\phi) \, e^{\frac{i}{\hbar} \delta S \cos(\phi)}
             \, d\phi\right\}.
\end{equation}
Here $A(\phi)$ is the Gutzwiller amplitude function in the trace formula for
the perturbed system, ${\cal J}(\phi)=\partial q/\partial\phi$ is the Jacobian
due to the variable mapping, and $\Phi_0 = S_0/\hbar - \sigma_0\pi/2$ is the
overall phase (including the action $S_0$ and the Maslov index $\sigma_0$) of
the level density in the unperturbed system with U(1) symmetry. The quantity
$\delta S \cos{(\phi)}$ in the exponent is the action shift caused by the
symmetry breaking term in the Hamiltonian. For a first inspection, perturbation
theory might give a hint to the value of the constant $\delta S$ (which depends
on the energy and on the parameters in the symmetry breaking term,
such as deformation, nonlinearity, etc.). 
Hereby one may have to go beyond the first order of the
perturbation expansion. Recent work on H\'enon-Heiles systems \cite{bcl} gives
us a hint that going to the lowest order at which $\delta S$ becomes nonzero --
however high it may be -- can be combined with keeping the unperturbed
amplitude $A_0$. Putting $A(\phi){\cal J}(\phi)=A_0$, which one may do in the
small-perturbation limit, the $\phi$ integral in \eq{U1int} can be done
analytically, and one obtains
\begin{equation} \label{U1pert}
 \delta g = A_0 J_0(\delta S/\hbar) \cos(\Phi_0) \, ,
\end{equation}
where $J_0(x)$ is a standard cylindrical Bessel function. When $\delta S$ is
zero, we have the trace formula in the symmetric limit (for one orbit family)
\begin{equation} \label{U1sym}
 \delta g_0 = A_0 \cos(\Phi_0) = A_0 \cos(S_0/\hbar - \sigma_0 \pi/2) \, .
\end{equation}
For large deviations from the U(1) symmetry, i.e., for $\delta S \gg \hbar$, we
can use the asymptotic expansion of $J_0(x)\sim \sqrt{(2/\pi|x|)}\cos(x-\pi/4)$
to find
\begin{equation} \label{U1asy}
 \delta g \sim A_0 \sqrt{\hbar/2\pi |\delta S|}
               \left[\cos(\Phi_0 + \delta S/\hbar - \pi/4)
                   + \cos(\Phi_0 - \delta S/\hbar + \pi/4) \right] .
\end{equation}
This corresponds to the two isolated orbits with the action shifts $\pm \delta
S$ and the corresponding corrections to the Maslov indices. Note that the two
terms above arise from a saddle-point approximation to the integral in
\eq{U1int} at the stationary points $\phi_0=0$ and $\phi_0=\pi$, respectively.
The action shifts and amplitudes in \eq{U1asy} will in general be correct only
in the small-perturbation limit.

We want to find a uniform approximation that for $\delta S \gg \hbar$ reaches
the correct Gutzwiller trace formula for the pair of isolated orbits
\begin{equation} \label{U1gutz}
 \delta g_G = A_u \cos(S_u/\hbar - \sigma_u \pi/2)
            + A_s \cos(S_s/\hbar - \sigma_s \pi/2)\, ,
\end{equation}
where the indices $s$ and $u$ refer to the stable and unstable orbits,
respectively. We first define the following quantities:
\begin{equation} \label{comb}
 {\bar A} = \frac12 (A_u+A_s)\, , \quad \Delta A = \frac12 (A_u - A_s) \, ,
 \quad
 {\bar S} = \frac12 (S_u+S_s)\, , \quad \Delta S = \frac12 (S_u - S_s) \, ,
\end{equation}
\begin{equation} \label{avphas}
    \bar \Phi = {\bar S}/\hbar - {\bar \sigma}\, , \quad
{\bar \sigma} = \frac12 (\sigma_u + \sigma_s) = \sigma_0 \, .
\end{equation}
We now make the following ansatz for the uniform approximation, which consists
of expanding the product $A(\phi) {\cal J}(\phi)$ up to two terms with suitably
chosen coefficients:
\begin{equation}\label{U1unint}
  \delta g_u = \sqrt{2\pi|\Delta S|/\hbar} \; \Re e
                  \left\{e^{i{\bar \Phi}} \frac{1}{2\pi} \int_0^{2\pi}
                  ({\bar A} + \Delta A \cos\phi ) \,
                  e^{\frac{i}{\hbar} \Delta S \cos(\phi)}
                  \, d\phi\right\} .
\end{equation}
For small perturbations, $\Delta S \sim \delta S$, and therefore \eq{U1unint}
will by construction lead to the correct symmetric limit \eq{U1sym}, since the
divergence of the Gutzwiller amplitudes in this limit is given by the first
factor in Eq.\ \eq{U1asy}. On the other hand, the coefficients in parentheses
under the integral in \eq{U1unint} have been chosen such that in the asymptotic
limit $|\Delta S|\gg \hbar$, the stationary-phase evaluation will lead to the
amplitudes ${\bar A} \pm \Delta A$ which are precisely the Gutzwiller
amplitudes $A_u$ and $A_s$, respectively, and thus give the form \eq{U1gutz}.

The integration in \eq{U1unint} can be done analytically, using
\begin{equation}\label{bes1int}
  {1\over{2\pi}} \int_0^{2\pi} d\phi \cos\phi \, e^{ix\cos\phi} = i J_1(x)\,,
\end{equation}
and leads to the Tomsovic-Grinberg-Ullmo (TGU) uniform approximation \cite{TGU},
in the compact form given by Sieber \cite{siel}, for the contribution of a pair 
of symmetry-broken isolated orbits to the trace formula:
\begin{equation}\label{trTGU}
 \delta g_u = \sqrt{2\pi|\Delta S|/\hbar} \left\{
               {\bar A}\, J_0(\Delta S/\hbar) \cos(\bar \Phi)
               - \Delta A\, J_1(\Delta S/\hbar) \sin(\bar \Phi)
               \right\} .
\end{equation}
Note that this formula holds for all generic non-integrable systems in two
dimensions that arise from an integrable system with U(1) symmetry through a
symmetry-breaking term in the Hamiltonian that is governed by a continuous
parameter. Particular examples are two-dimensional billiards obtained by
deforming the circular billiard. The nature of the deformation parameter
generally plays no role. The only assumption made is that the original orbit
families (i.e., polygons in the case of the circular billiard) are broken into
pairs of stable and unstable isolated orbits. The modification that becomes
necessary when extra degeneracies due to discrete symmetries are present
is trivial and will be dealt with explicitly in the examples discussed below.
The breaking into orbit pairs is the most frequent situation. Exceptions 
occur, e.g., in billiards with octupole or hexadecapole deformations, where 
the boundary in polar coordinates is given by $r(\theta) = R\,[1+\epsilon_\ell 
P_\ell(\cos\theta)]$ with $\ell=3$ or 4. There the diameter orbit family 
breaks up into more than two isolated librating orbits (not counting 
discrete degeneracies) \cite{mesb}. These have to be treated with 
different (and more complicated) uniform approximations.

We also note that the deformation away from the integrable case should be small
enough that no bifurcation of the stable isolated orbits has taken place or is
about to arise. Near the bifurcation points, the known uniform approximations
apply \cite{rich,bif,siel} which we shall not discuss here.

\section{Uniform approximations for SU(2) breaking}

No uniform approximation has, to our knowledge, been derived so far for
systems with higher than U(1) symmetry. In the following, we shall do so for
two systems obtained by breaking the SU(2) symmetry of the two-dimensional
harmonic oscillator. We shall follow the heuristic way of deriving the
uniform approximation described in the previous section, starting from the
perturbative limit which here is treated using the approach of Creagh
\cite{crp}.

For isotropic and anisotropic harmonic oscillators in any dimension, analytical
trace formulae are known which converge to the exact quantum-mechanical sum of
delta functions \cite{book,bj}. For the two-dimensional isotropic case, the
oscillating part of the level density is
\begin{equation}\label{ho2}
  \delta g_0 = A_0 \sum_{r=1}^\infty \cos\left(\frac{rS_0}{\hbar}\right),
  \qquad A_0 = \frac{2E}{(\hom)^2}, \qquad S_0 = \frac{2\pi E}{\omega}.
\end{equation}
(Note that the smooth Thomas-Fermi part is given by $A_0/2$.) As pointed out in
\cite{crp}, the continuous degeneracy of the classical periodic orbits in this
system can be described by integration of the surface element $d\Omega =
\sin\beta\, d\beta\, d\gamma$ of a unit sphere:
\begin{equation}\label{su2int}
  \frac{1}{4\pi}  \int_0^{2\pi} d\gamma \int_0^\pi \sin\beta \, d\beta = 1\, .
\end{equation}
As a result of the SU(2) symmetry, the action $S_0$ is independent of the
angles $\beta$ and $\gamma$. In the presence of a small perturbation, the
periodic orbits will be distorted, resulting in an action shift $\delta S$ that
in general will depend on $\beta$ and $\gamma$. Explicit ways of calculating
$\delta S(\beta,\gamma)$ starting from a Hamiltonian of the form \eq{Hpert} are
given in \cite{crp}. For small values of the perturbation parameter $\epsilon$,
the main effects governing the level density will come from the action shift in
the phase of the trace formula, whereas the unperturbed amplitude $A_0$ can be
retained. The perturbative trace formula for symmetry breaking then reads as
\begin{equation}\label{dgpert}
   \delta g_{pert} = A_0 \, \Re e \left\{ \sum_{r=1}^{r_m} \M(rx)\,
                     e^{irS_0/\hbar} \right\}.
\end{equation}
Here the modulation factor $\M(x)$ (which in general is complex) is given by
the average of the phase shift, taken over the originally degenerate periodic
orbit family,
\begin{equation}\label{MSU2}
 \M(x) = \frac{1}{4\pi}  \int_0^{2\pi} d\gamma \int_0^\pi \sin\beta \, d\beta
         \, e^{i\delta S(\beta,\gamma)/\hbar},
\end{equation}
and $r$ is the repetition number. The dimensionless quantity $x$ is
proportional to $\epsilon$ and inversely proportional to $\hbar$, and depends
on some power of the energy $E$. For $\epsilon\rightarrow 0$ and hence
$x\rightarrow 0$, we have $\delta S\rightarrow 0$ so that $\M\rightarrow 1$,
and the unperturbed trace formula \eq{ho2} is recovered. The repetition number
$r$ in \eq{dgpert} cannot be summed up to arbitrarily high values, since the
argument $rx$ must remain of order unity or smaller for the perturbation
approach to be valid. Hence, the maximum value $r_m$ must be chosen such that
$r_m x \siml 1$ for given values of $\epsilon$ and $E$. This excludes in
general the possibility of quantizing the system through the trace formula in
this approach. However, we shall be interested only in the low-frequency
components of the oscillating level density, i.e., in its gross-shell structure
that is governed by the shortest periodic orbits and their first few
repetitions \cite{book,stma}.

Our main task now is to generalize the modulation factor \eq{MSU2} in such a
way that the trace formula \eq{dgpert} goes over to the Gutzwiller formula
\cite{gutz} for isolated orbits in the limit of large perturbations $\epsilon$
that fully break the symmetry, whereas the limit $\M(0)=1$ is preserved. If we
succeed in finding such a generalization, it will smoothly interpolate between
the exact trace formula \eq{ho2} for the harmonic oscillator and the Gutzwiller
formula for the symmetry-broken limit, and hence be a suitable uniform
approximation. Note that Eq.\ \eq{dgpert} with \eq{MSU2} is exactly of the same
form as Eq.\ \eq{U1int} for the U(1) case, except that we now have a two-fold
integral. We can therefore take the same course of action to find a uniform
approximation that has the above two limits: ($i$) evaluate the asymptotic
amplitudes for large values of $x$ (i.e., 
for large $\epsilon$), ($ii$) map the exact
action shift $\delta S$ onto the form obtained from perturbation considerations
(but with freely adjustable parameters), ($iii$) include under the integral a
parameterized expression of the same form for the product of the amplitude
function $A$ and the Jacobian $\cal J$ of the mapping function, and ($iv$)
adjust all the parameters such that in the asymptotic limit of large $x$ the
correct Gutzwiller amplitudes and actions of the isolated orbits are obtained,
while the unperturbed limit \eq{ho2} is kept for $x=0$.

This procedure is very similar in spirit to that used by Sieber and Schomerus
in their uniform treatment of many types of bifurcations \cite{bif}. However,
instead of starting from a phase-space representation of the trace formula and
expanding the action in normal forms, we use the group integral representation
of the (unperturbed) trace formula developed by Creagh and Littlejohn
\cite{gsym}. The latter exploits directly the properties of the symmetry group
characteristic of the system $H_0$; it has been used as a starting point for
the perturbative trace formula of Creagh \cite{crp}, as shown above in Eqs.\
\eqq{dgpert}{MSU2}.

Different from the U(1) case (with the exceptions mentioned at the end of the
previous section), the explicit results which we obtain here do depend on the
explicit form of the Hamiltonian $H_1$. On the other hand, one of the results
(that for the standard H\'enon-Heiles potential) turns out to apply also to a
three-dimensional cavity with small axially-symmetric quadrupole deformations.
This is due to the close relation between the SU(2) and the SO(3) symmetry that
is broken in the latter case. We will discuss this system in section IV.


\subsection{The quartic H\'enon-Heiles potential}

We first investigate the quartic H\'enon-Heiles (HH4) potential in two
dimensions, given in polar coordinates by
\begin{equation}\label{Vhh4}
  V(r,\theta) = \frac{1}{2}\,\omega^2 r^2-{\alpha\over 4}\, r^4\cos(4\theta)\,,
\end{equation}
which has recently been investigated \cite{bcl} in the framework of the
semiclassical perturbation theory described above. The limit $\alpha=0$ is a
harmonic oscillator with the SU(2) symmetry. The anharmonic term makes the
system non-integrable with mixed classical dynamics. It retains a discrete
four-fold rotational symmetry with four saddle-points at the energy $E^* =
\omega^4/4\alpha$, through which the particle can escape. The shortest periodic
orbits in this system are two pairs of straight-line librating orbits (named
A$_1$ and A$_2$ in Ref.\ \cite{bcl}) and a circulating orbit (named C). Like in
the standard cubic H\'enon-Heiles (HH) potential \cite{bcl,hh,bblm}, the system
is scaled with $\alpha$ and its dynamics can be described in terms of one
continuous parameter, the scaled energy $e$ defined by
\be \label{H4scal}
e=E/E^*= 4\alpha E/\omega^4.
\ee
The actions of the periodic orbits are changed in first order of the
perturbation by a shift
\begin{equation}\label{S1hh4}
   \delta S_1 = \hbar x \, (\sin^2\!\beta\,\cos^2\!\gamma - \cos^2\!\beta)\,,
   \qquad  \hbar x = S_0\, \frac{3}{32}\, e\,.
\end{equation}
Then, the perturbative modulation factor $\M(x)$ in \eq{MSU2} becomes (with
$u=\cos\beta$)
\begin{equation}\label{M441}
 \M(x) = \frac{1}{4\pi} \int_0^{2\pi} d\gamma \int_0^\pi \sin\beta \, d\beta\,
         e^{ix\, (\sin^2\!\beta\,\cos^2\!\gamma - \cos^2\!\beta)}
       = \frac{1}{2\pi} \int_0^{2\pi} d\gamma \int_0^1 du\,
         e^{ix\, [(1-u^2)\cos^2\!\gamma - u^2]}.
\end{equation}
An alternative expression of this is obtained by rotating the unit sphere about
an angle $\pi/4$ along the 2 axis:
\be \label{rot}
{\bf e}_1'  =  {1\over\sqrt{2}}\,{\bf e}_1 - {1\over\sqrt{2}}\,{\bf e}_3\,,
                                                                  \quad
{\bf e}_2'  =  {\bf e}_2\,,                                       \quad
{\bf e}_3'  =  {1\over\sqrt{2}}\,{\bf e}_1 + {1\over\sqrt{2}}\,{\bf e}_3\,,
\ee
which leads to
\begin{equation}\label{M442}
 \M(x) = \frac{1}{4\pi} \int_0^{2\pi} d\gamma \,
         \int_0^\pi \sin\beta \, d\beta \,
         e^{ix\, \sin(2\beta)\,\cos\gamma}.
\end{equation}
This integral can be performed analytically (see \cite{bcl}) and yields
\be\label{M44}
   \M(x) = {\pi\over{2\sqrt{2}}} \; J_{-1/4}\left({x\over 2}\right)
           J_{1/4}\left({x\over 2}\right).
\ee
The asymptotic expansion of the Bessel functions $J_{\mu}(x)$ for large $x$
leads to three terms which we label according to the names of the periodic
orbits:
\begin{equation}\label{M44asy}
  \M(x)   \sim  \M_C(x) +  \M_{A_1}(x) + \M_{A_2}(x)\,, \qquad \qquad (x\gg 1)
\end{equation}
with
\bea\label{44orbs}
\M_C(x)     = \frac{1}{2x}  \qquad \qquad \qquad \qquad \quad \quad \;
    & \Rightarrow & \qquad   \, A_C \sim \frac{A_0}{2x},      \nonumber \\
\M_{A_1}(x) = \frac{1}{2\sqrt{2}\,x}\, \cos(+x-\pi/2)      \qquad
    & \Rightarrow & \qquad A_{A_1} \sim \frac{A_0}{2\sqrt{2}\,x}, \nonumber \\
\M_{A_2}(x) = \frac{1}{2\sqrt{2}\,x}\, \cos(-x+\pi/2)      \qquad
    & \Rightarrow & \qquad A_{A_2} \sim \frac{A_0}{2\sqrt{2}\,x}.
\eea
The first term corresponds to the loop orbit C, the second and third terms to
straight-line libration orbits A$_1$ and A$_2$, respectively; each of them has
a discrete degeneracy of two which is included in the amplitudes. These are
\cite{bcl} the only periodic orbits with periods of order $T_0=2\pi/\omega$ 
up to energies $e\siml 0.85$. The forms on
the left-hand side of Eq.\ \eq{44orbs} contain the exact Maslov indices
$\sigma_C=0$, $\sigma_{A_1}=+1$, and $\sigma_{A_2}=-1$ of the isolated orbits
and, with Eq.\ \eq{ho2}, yield the asymptotic amplitudes of the trace formula
shown on the right-hand side of Eq.\ \eq{44orbs}. For small $x$, these
amplitudes have been shown in Ref.\ \cite{bcl} to reproduce numerically well
the diverging Gutzwiller amplitudes of the isolated periodic orbits. Note that
the action shifts predicted at the first order of the perturbation theory are
zero for the orbit C and $\pm \hbar x$ for the orbits A$_1$ and A$_2$,
respectively. This can be checked numerically for all orbits, as well as
analytically for the orbits A$_1$ and A$_2$. The actions of the latter, being
straight-line one-dimensional integrals, can be expressed analytically in terms
of complete elliptic integrals (see the appendix) and then be Taylor expanded
in powers of $e$. The result is
\begin{eqnarray} \label{SAexp}
  S_{A_1} & = & S_0\left(1+\frac{3}{32}e+\frac{35}{1024}e^2
                          +\frac{1155}{65536}e^3+\dots\right),    \nonumber \\
  S_{A_2} & = & S_0\left(1-\frac{3}{32}e+\frac{35}{1024}e^2
                          -\frac{1155}{65536}e^3+\dots\right),
\eea
which confirms the first-order action shifts $\pm\hbar x$ given in Eqs.\
\eqq{S1hh4}{44orbs}.

For the following, it is important to trace back the asymptotic forms
\eq{44orbs} to singular points of the integral \eq{M441}. For this, we first
evaluate the asymptotic form of the $u$ integral. The stationary point at $u=0$
yields a term $\sim 1/\!\sqrt{x}$, and the end point at $u=1$ yields a term
$\sim 1/x$. The latter, after an exact integration over $\gamma$, gives the
contribution $\M_{A_2}(x)$. The $\gamma$ integral over the first term has four
stationary points; saddle-point integration at $\gamma=0$ and $\pi$ yields the
contribution $\M_{A_1}(x)$, and at $\gamma=\pi/2$ and $3\pi/2$ it yields the
contribution $\M_C(x)$. In summary, we get the periodic orbit contributions
asymptotically as follows:
\bea
\hbox{orbit A}_2: & \qquad \hbox{from } u=1, & \quad (\hbox{any }\gamma)\,,
                                                                   \nonumber \\
\hbox{orbit A}_1: & \qquad \hbox{from } u=0, & \quad \gamma=0 \hbox{ and }\pi\,,
                                                                   \nonumber \\
\hbox{orbit C}:   & \qquad \hbox{from } u=0, & \quad \gamma=\pi/2 \hbox{ and }
                                                                      3\pi/2\,.
\eea

We now construct a uniform approximation which for large $x$ yields the correct
asymptotic Gutz\-wil\-ler trace formula with the contributions from the three
leading isolated orbits A$_1$, A$_2$, and C. For describing the gross-shell
structure of the level density, it is sufficient to include only the lowest few
harmonics,\footnote{A more consistent truncation of the trace formula, followed
below, is achieved by Gaussian smoothing over a small energy range $\gamma$,
resulting in the amplitudes multiplied by exponential damping factors; see
Ref.\ \cite{bblm} for details. These exponential factors are included in the
amplitudes $A_{A_1}$, $A_{A_2}$, and $A_C$ and their combinations defined
below.} i.e., the first few repetitions $r$ of the primitive orbits, in the
trace formula. In the following, we shall only give the results for $r=1$;
higher repetitions can be included according to Eq.\ \eq{dgpert}. Hence, the
Gutzwiller limit is written as
\begin{equation}\label{dgr4gutz}
  \delta g_G =
       A_{A_1} \cos\left(\frac{S_{A_1}}{\hbar}-\sigma_{A_1}\frac{\pi}{2}\right)
     + A_{A_2} \cos\left(\frac{S_{A_2}}{\hbar}-\sigma_{A_2}\frac{\pi}{2}\right)
     + A_C     \cos\left(\frac{S_C}{\hbar}-\sigma_C \frac{\pi}{2}\right)\,,
\end{equation}
This limit can be imposed by including under the integral \eq{M441} the product
of amplitude function $A(u,\gamma)$ times Jacobian ${\cal J}(u,\gamma)$ in the
same form as in the exponent, but with different parameters:
\begin{eqnarray}\label{M44uni}
 \M_u(x) & = & \frac{1}{2\pi} \int_0^{2\pi} d\gamma \int_0^1 du\,
               e^{ix\, [(1-u^2)\cos^2\!\gamma - u^2]} \times
               \qquad \qquad \qquad           \nonumber \\
         &   & \qquad \qquad \qquad \qquad \times
               \frac{2x}{A_0} \left\{   \sqrt{2} A_{A_2} u^2 +
               (1-u^2)\left[\sqrt{2} A_{A_1} \cos^2\!\gamma + A_C \sin^2\!\gamma
                      \right] \right\}.
\end{eqnarray}
Note that the coefficients in the second line are precisely the inverses of the
asymptotic amplitudes given in \eq{44orbs}. In this way, the approximation
\eq{M44uni} leads by construction to the Gutzwiller limit \eq{dgr4gutz} for
large $x$, whereas for $x\rightarrow 0$ the diverging Gutzwiller amplitudes
exactly cancel altogether and $\M_u\rightarrow 1$, as required.

The integrals occurring in \eq{M44uni} can all be done analytically. For the
integral appearing as the coefficient of $A_C$, we note that ${\bf e}_2^2 =
\sin^2\!\beta \sin^2\!\gamma$ is invariant under the rotation \eq{rot}, so that
we can replace the phase in the exponent by that appearing in Eq.\ \eq{M442}.
Then, using the same transformations as in Ref.\ \cite{bcl} for obtaining the
form \eq{M44} of the perturbative modulation factor, and exploiting some
recurrence relations amongst the Bessel functions $J_\mu(x)$, we obtain
\bea \label{intAC}
 \frac{1}{2\pi} \int_0^{2\pi} d\gamma \,\frac12 \int_0^\pi \sin\beta \, d\beta\,
 \sin^2\!\beta\sin^2\!\gamma\,e^{ix\, \sin(2\beta)\,\cos\gamma}
\qquad \qquad \qquad \qquad \qquad \qquad \qquad    \nonumber \\
\qquad \qquad \qquad \qquad \qquad
         = {\pi\over{4\sqrt{2}}}\,\left[ J_{-1/4}\left({x\over 2}\right)
                                 J_{1/4}\left({x\over 2}\right)
                               - J_{-3/4}\left({x\over 2}\right)
                                 J_{3/4}\left({x\over 2}\right) \right].
\eea
The other two integrals can be found first by taking the derivative of Eq.\
\eq{M441} with respect to $x$, and second by integrating the terms proportional
to $u^2$ in \eq{M44uni} over $u$ by parts, and finally by taking suitable
linear combinations of the results of these two operations. The final
expression is
\be \label{M44unires}
 \M_u(x) = \frac{2x}{A_0} \left[ A_C\M_{-}(x) + \sqrt{2} {\bar A}_A \M_{+}(x)
           -i\sqrt{2}\, \Delta\! A_A \M'(x) \right],
\ee
where
\be
{\bar A}_A = \left(\frac{A_{A_1}+A_{A_2}}{2}\right), \qquad
\Delta\! A_A = \left(\frac{A_{A_1}-A_{A_2}}{2}\right),
\ee
and
\bea \label{M44pm}
\M_{\pm}(x) & = & {\pi\over{4\sqrt{2}}}\,\left[ J_{-1/4}\left({x\over 2}\right)
                                        J_{1/4}\left({x\over 2}\right)
                                   \pm J_{-3/4}\left({x\over 2}\right)
                                        J_{3/4}\left({x\over 2}\right) \right],
                                                        \nonumber \\
   \M'(x) & = & - {\pi\over{4\sqrt{2}}}\,\left[ J_{-1/4}\left({x\over 2}\right)
                                        J_{5/4}\left({x\over 2}\right)
                                      + J_{1/4}\left({x\over 2}\right)
                                        J_{3/4}\left({x\over 2}\right) \right].
\eea

The last step now is to insert the modulation factor \eq{M44unires} into Eq.\
\eq{dgpert} and to replace the unperturbed action $S_0$ in the phase of the
first term of \eq{M44unires} by $S_C$, and for the other two terms by the
average action ${\bar S}_A$ of the orbits $S_{A_1}$ and $S_{A_2}$, while the
perturbative action shift $\hbar x$ is redefined as their difference $\Delta
S_A$
\be
\hbar x = \Delta S_A = \frac12 \, (S_{A_1}-S_{A_2})\,, \qquad
{\bar S}_A = \frac12 \, (S_{A_1}+S_{A_2})\,.
\ee
By these replacements, we ensure that the phases in the asymptotic level density
\eq{dgr4gutz} contain the correct numerical actions of the isolated orbits.

The final form of the uniform level density for the HH4 potential (including
only the primitive orbits, i.e., $r=1$) is then:
\bea \label{dghh4u}
\delta g_u & = & (2\Delta S_A/\hbar) \, \Re e \left\{
                         e^{\frac{i}{\hbar}S_C}
                         A_C \M_{-}\left(\Delta S_A/\hbar\right) \right.
                                                                 \nonumber \\
           &   & \left. \qquad \qquad
                       + \sqrt{2}\, e^{\frac{i}{\hbar}{\bar S}_A}
                         \left[ {\bar A}_A
                         \M_{+}\left(\Delta S_A/\hbar\right)
                       - i\, \Delta\! A_A
                       \M'\left(\Delta S_A/\hbar\right) \right] \right\}
                                                                 \nonumber \\
           & = & (2\Delta S_A/\hbar) \left\{ A_C \M_{-}             
                 \left(\Delta S_A/\hbar\right)\,\cos\!\left(S_C/\hbar\right)
                 \right.                                   \nonumber \\
           &   & \left. \qquad \qquad
               + \sqrt{2} \left[ {\bar A}_A \M_{+}
                 \left(\Delta S_A/\hbar\right)\,
                 \cos\!\left({\bar S}_A/\hbar\right)
               + \Delta\! A_A \M'\left(\Delta S_A/\hbar\right)\,
                 \sin\!\left({\bar S}_A/\hbar\right)
                 \right] \right\}\!.
\eea

As in the case of the TGU uniform approximation discussed in Sec.\ II, this
formula only 
holds as long as no bifurcations of the stable isolated orbits occur. 
For the primitive orbits of the HH4 system, this does not happen up to
$e\simeq 0.85$, where the primitive orbit A$_1$ undergoes an isochronous
bifurcation. In the numerical results shown below, we have smoothed the
Gutzwiller amplitude of this orbit (see Ref.\ \cite{bcl} for details) in order
to simulate a better treatment of this bifurcation by the corresponding
uniform approximation \cite{bif}.

In \fig{r4compps} we show a numerical compilation of the three semiclassical
approximations to the level density $\delta g$ discussed here (given by dashed
lines) and compare them to the quantum-mechanical one (given by the solid
lines). In all cases, Gaussian averaging over the energy $E$ with a range
$\gamma=0.6 \,\hom$ has been applied. This damps the amplitudes strongly enough
so that only the primitive orbits ($r=1$) need be kept in the semiclassical
approaches. At the bottom of this figure, the Gutzwiller result \eq{dgr4gutz}
for the isolated orbits is shown. It gives an excellent agreement with the
quantum result for $E\simg 10 \,\hom$, even up to $e\simeq 1$ (the saddle energy
is $E^*\simeq 39\,\hom$ for the case $\alpha=0.0064$ chosen here). For small
energies, it diverges due to the approaching SU(2) symmetry limit. In the
middle of the figure, the perturbative result \cite{bcl} is shown. It
reproduces the quantum result up to $E\sim 13\,\hom$, thus catching the essential
feature of the symmetry breaking. (Note that at $E\sim 13\,\hom$ we have $x\sim
3$, showing that the perturbative approach may be used also for values of $x$
somewhat larger than unity, cf.\ \cite{crp,bcl}.) For $E>14\,\hom$ the
anharmonicity is, however too large for perturbation theory to apply. Finally,
the top part of \fig{r4compps} shows the present uniform approximation that
reproduces the quantum result at all energies.

\Figurebb{r4compps}{60}{28}{767}{363}{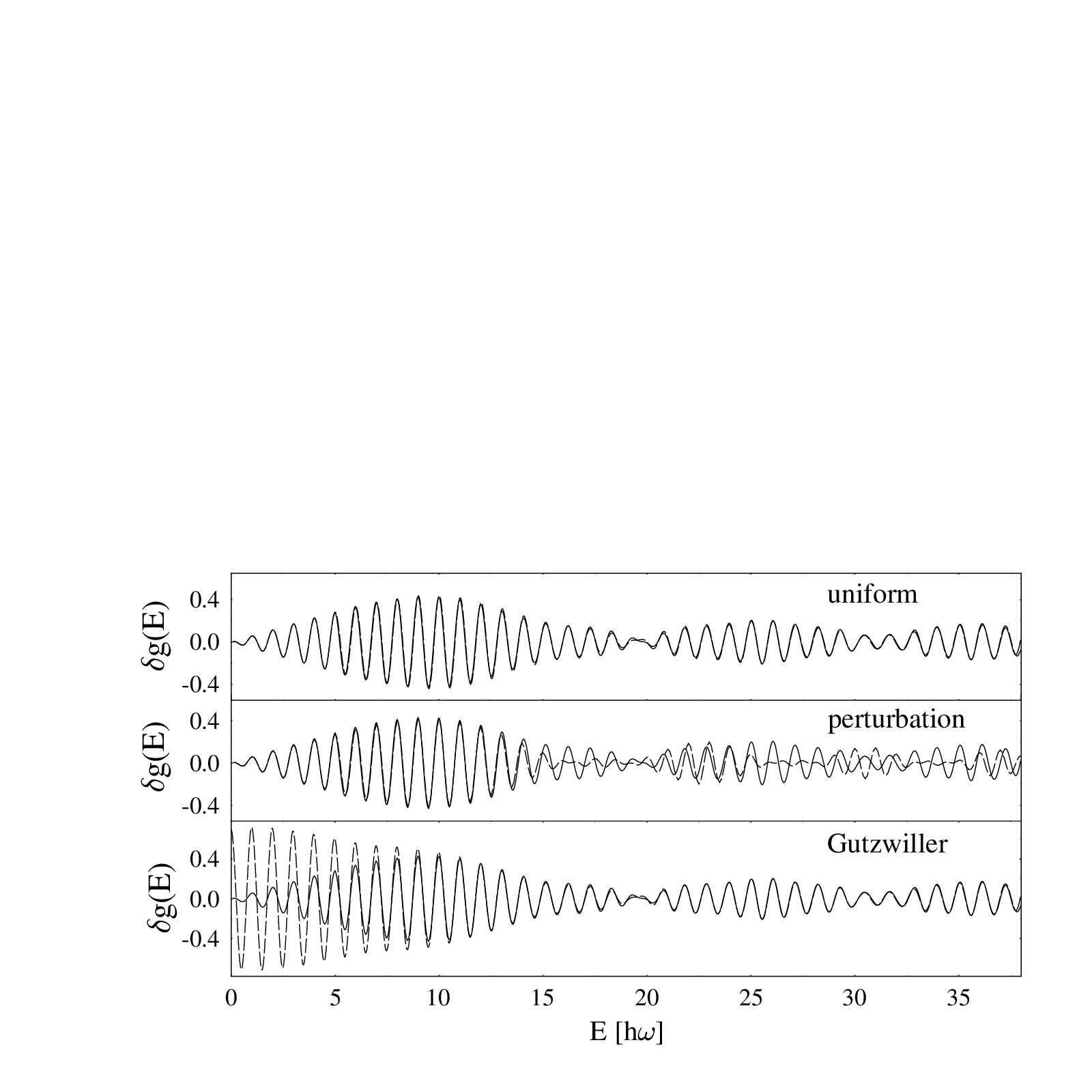}{7.2}{15.0}{
Oscillating part of level density of the HH4 potential ($\alpha=0.0064$),
averaged over a range $\gamma=0.6$, versus energy $E$ (units: $\hom$). 
Solid lines: quantum-mechanical results. Dashed lines: semiclassical
approximations (with $r_m=1$). Bottom: Gutzwiller trace formula
(diverging at small energies); middle: perturbative trace formula
by Creagh (failing at large energies); top: present uniform
approximation (working at all energies).
}

In \fig{r4udgps} we compare the results of the present uniform approximation
(dashed lines) with the quantum results (solid lines) for two different values
of the anharmonicity parameter $\alpha$. In the two upper panels, the same
smoothing width $\gamma=0.6\,\hom$ was used as in the previous figure and only
the primitive orbits ($r=1$) A$_1$, A$_2$ and C were included. In the two lower
panels, we have used a smaller smoothing width $\gamma=0.25\,\hom$ and included
the second repetitions (i.e., $r_m=2$) of the three orbits as well. Now a more
detailed fine structure of the level spectrum is resolved; even this is well
reproduced by the semiclassical approximation. Only in the regions
corresponding to $e=E/E^*\simg 0.8$ ($E\simg 31 \,\hom$), some small differences
can be seen which are explained by the fact that at these energies, new orbits
with actions comparable to those of the included orbits with $r=2$ exist (after
period-doubling bifurcations of the orbits C and A$_1$) which we have not
included. Their inclusion would necessitate a proper uniform treatment of the
corresponding bifurcations, which is outside the scope of this paper.

In the actual computation of the uniform trace formula for small energies $e$,
cancellations between the diverging Gutzwiller amplitudes take place. This
requires their rather accurate numerical determination. For practical purposes,
it is advisable to take advantage of the fact that for sufficiently small
arguments $x=\Delta S_A/\hbar$, the result \eq{dghh4u} goes over into the
perturbative trace formula \eq{dgpert} with the modulation factor \eq{M44}
which is numerically much more robust. We found that one may switch between the
two formulae for values $1.5 \siml x \siml 2.5$ without visibly changing the
results shown above.

\Figurebb{r4udgps}{85}{78}{795}{512}{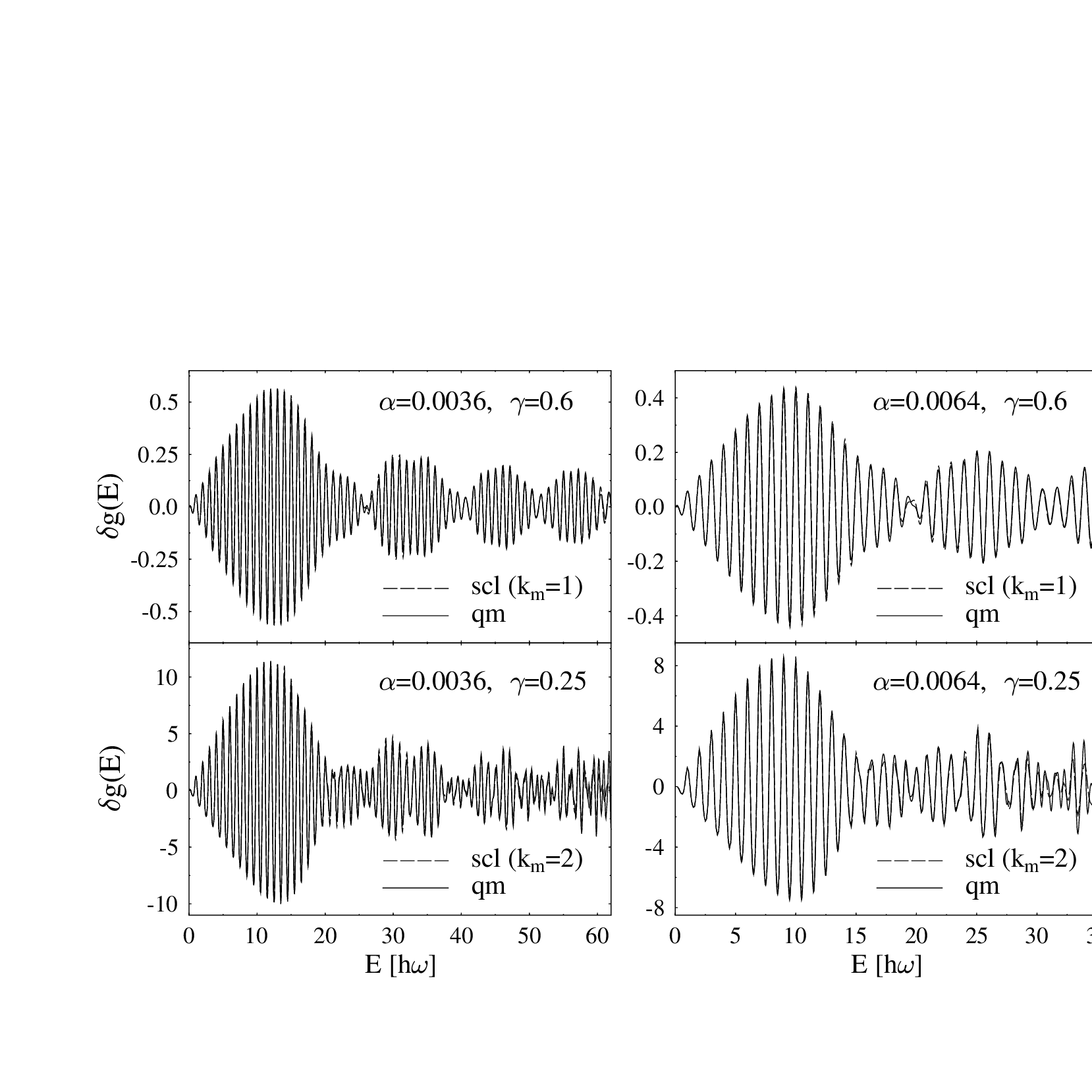}{10.0}{16.0}{
Oscillating part of level density of the HH4 potential versus energy $E$
(units: $\hom$). Solid lines: quantum-mechanical results. Dashed lines:
semiclassical results with present uniform approximation. Left side:
$\alpha=0.0036$ ($E^*\simeq 69\,\hom$), right side: $\alpha=0.0064$
($E^*\simeq 39\,\hom$). Upper part: $\gamma=0.6$ and $r_m=1$, lower part:
$\gamma=0.25$ and $r_m=2$.
}


\subsection{The standard H\'enon-Heiles potential}

We will now investigate the original H\'enon-Heiles (HH) potential \cite{hh},
given in polar coordinates by
\begin{equation}\label{Vhh}
  V(r,\theta) = \frac{1}{2}\,\omega^2 r^2-{\alpha\over 3}\, r^3\cos(3\theta)\,.
\end{equation}
The three shortest isolated orbits in this potential, which have been shown
\cite{bblm} to govern the beating gross-shell structure in the level density of
this system, are a straight-line librating orbit A, a curved librating orbit B,
and a circulating orbit C (similar to that in the HH4 potential). Orbits A and
B have a discrete degeneracy of three, due to the three-fold discrete rotational
symmetry of the potential \eq{Vhh}, and the orbit C is two-fold degenerate due
to time reversal symmetry.

Because of the odd power of the anharmonic term in \eq{Vhh}, the first-order
perturbation result is zero both classically and quantum-mechanically. In
second-order perturbation, the action shift $\delta S$ due to the anharmonicity
is given \cite{bcl} by
\begin{equation}\label{S2hh}
   \delta S_2 = \hbar x \, (5 - 7 \cos^2\!\beta )/6\,,
\end{equation}
with
\begin{equation}
   \hbar x = \frac{S_0}{12} \frac{\!E}{E^*} = \frac{S_0}{12} e \,, \quad
   E^*=\frac{\omega^6}{6\alpha^2}\,.
\end{equation}
Here $E^*$ is the saddle energy for the potential \eq{Vhh}, and $S_0$ is the
action of the unperturbed harmonic oscillator given in \eq{ho2}. The
perturbative modulation factor $\M(x)$ becomes
\begin{equation}\label{Mhhint}
   \M(x) = {1\over {4\pi}} \int d\Omega \;e^{ix(5-7u^2)/6}
         = \int_0^1 du \; e^{ix(5-7u^2)/6}.
\end{equation}
This integral can be expressed analytically in terms of the Fresnel functions
$C(x)$ and $S(x)$ (we use the convention of Abramowitz and Stegun \cite{abst}):
\begin{equation}\label{Mhh}
   \M(x) = e^{5ix/6} \; {1\over\xi}
           \left[ C(\xi)-iS(\xi) \right] ,
           \qquad \xi=\sqrt{ {{7|x|}\over{3\pi}} }\,.
\end{equation}
Using the asymptotic expansion of the Fresnel functions for large arguments
$x \gg 1$,
\begin{eqnarray}\label{CSasy}
  C(\xi) & \sim & \frac 12 + \frac{1}{\pi \xi} \sin(\pi \xi^2/2) \,,
  \nonumber \\
  S(\xi) & \sim & \frac 12 - \frac{1}{\pi \xi} \cos(\pi \xi^2/2) \,,
\end{eqnarray}
we get the asymptotic result
\begin{equation}\label{dghhasy}
  \delta g \sim \sqrt{\frac{3\pi}{14|x|}}A_0\cos(S_0/\hbar+5x/6-\pi/4)
           + \frac{3}{7|x|} A_0 \cos(S_0/\hbar-x/3+\pi/2)\,,
\end{equation}
which is again only correct for sufficiently small values of the scaled energy
$e$ so that $x \siml 1$. Note that \eq{dghhasy} predicts the action shift
$\Delta S_C$ and the average action shift $\Delta S_{AB}$ to be
\begin{equation}
\Delta S_C = S_C - S_0 = -\hbar x / 3\,, \quad
\Delta S_{AB} = {\bar S}_{AB} - S_0 = 5 \hbar x / 6\,, \quad
{\bar S}_{AB} = \frac12 (S_A+S_B)\,.
\label{hhactdif2}
\end{equation}
This is numerically well fulfilled at lower energies. The expansion
\eq{dghhasy} is also obtained from the asymptotic analysis of the integral
\eq{Mhhint}, whereby the first term comes from the stationary point at $u=0$
and the second term is the end-point correction from $u=1$. The latter
corresponds to the C orbit in the HH potential with the correct Maslov index
\cite{bcl} $\sigma_C=-1$. Its amplitude is thus predicted in the second-order
perturbation theory as
\be  \label{ACasy}
 A_C^{pt2} = \frac{6E}{7(\hom)^2x} = \frac{36}{7\pi\hom}\,\frac{1}{e},
\ee
which was shown in Ref.\ \cite{bcl} to describe well the numerical Gutzwiller
amplitude $A_C$, not only for small $e$ where it diverges, but also up to
energies $e\simeq 1$. The first term in \eq{dghhasy} corresponds to the sum of
orbits A and B which, however, are still degenerate on the circle $\gamma \in
[0,2\pi)$ at the second order of the perturbation expansion and therefore have
an asymptotic amplitude too large by a factor $1/\sqrt{\hbar}$. These orbits
are broken up only at fourth order, with an action shift
\begin{equation}\label{S4hh}
   \delta S_4 = - \hbar y \sin^3\beta \cos(3\gamma)\,, \quad  y = c_4 e^3 .
\end{equation}
Including this into the phase of the modulation factor, we get
\begin{eqnarray}\label{Mhhint2}
   \M(x,y) & = & \int_0^1 du \, e^{ix(5-7u^2)/6}\,
               {1\over {2\pi}} \int_0^{2\pi} \!\! d\gamma \,
               e^{-iy(1-u^2)^{3/2} \cos(3\gamma)} \nonumber \\
           & = & \int_0^1 du \, e^{ix(5-7u^2)/6}\, J_0[y(1-u^2)^{3/2}]\,.
\end{eqnarray}
In the asymptotic expansion of this modulation factor, the Bessel function
modifies only the contribution from the stationary point at $u=0$,
leading to
\begin{equation}\label{dghhasy2}
  \delta g \sim \sqrt{\frac{3\pi}{14|x|}}A_0 J_0(y) \cos(S_0/\hbar+5x/6-\pi/4)
           + \frac{3}{7|x|} A_0 \cos(S_0/\hbar-x/3+\pi/2)\,.
\end{equation}
Upon further expansion of $J_0(y)$ for $y \gg 1$, this yields the amplitudes and
the correct Maslov phases \cite{bcl} $\sigma_A=1$ and $\sigma_B = 0$ of the
(now isolated) orbits A and B. The amplitudes, which are equal at this order
of the perturbation theory, go like $1/\sqrt{|xy|}$ and thus have the same
power of $\hbar$ as $A_C^{pt2}$. The actions of the orbits A and B are shifted
from the average value ${\bar S}_{AB}$ by an amount $\pm\hbar y$, respectively.
For the orbit A, this can again be checked by analytical integration (see the
appendix) and Taylor expansion of its period. The action becomes
\be \label{SAHexp}
  S_A = S_0\left(1+\frac{5}{72}e+\frac{385}{15552}e^2
                  +\frac{85085}{6718464}e^3+\dots\right).
\ee
The first correction term is the average action shift $\Delta S_{AB}$
\eq{hhactdif2}, obtained for both orbits A and B at second order of the
perturbation theory; the next term gives the value of $c_4$ in Eq.\ \eq{S4hh}.
With this, the fourth-order prediction of the average amplitude of orbits A
and B becomes
\be  \label{AABasy}
 A_{AB}^{pt4} = \frac{E}{(\hom)^2}\sqrt{\frac{3}{7|xy|}}
              = \frac{216}{7\pi\hom}\sqrt{\frac{3}{55}}\,\frac{1}{e^{3/2}}.
\ee
[Note that this amplitude contains the degeneracy factor 3 of the orbits A and
B mentioned above; it is related to the factor 3 in the argument of the cos
function in \eq{S4hh}. Similarly, $A_C^{pt2}$ \eq{ACasy} contains the
time-reversal factor 2 of the orbit C, which is related to the two end-point
corrections, coming from the $u$ integral in \eq{Mhhint} which originally runs
from $-1$ to $+1$, each giving one half of the second term in \eq{dghhasy}.]

We now redefine the quantities $x$ and $y$ in terms of the true actions of the
isolated periodic orbits
\begin{equation}
\hbar y  =  \delta S = \frac12 (S_A-S_B)\,,\quad
\frac76 \hbar x  = \Delta S =  \frac12 (S_A + S_B) - S_C\,,
\label{hh3actdif}
\end{equation}
and introduce the combinations of the Gutzwiller amplitudes
\begin{equation}
   {\bar A}_{AB} = \frac12 (A_A + A_B)\,, \quad
   \Delta A_{AB} = \frac12 (A_A - A_B)\,.
\end{equation}
The uniform approximation to the modulation factor is then
\begin{eqnarray}\label{Mhhuni}
 \M_u & = & \frac{1}{A_0}\! \int_0^1 du \, e^{(i/\hbar)\,\Delta S\,(5/7-u^2)}\,
            {1\over {2\pi}}\! \int_0^{2\pi} \!\! d\gamma\,
            e^{-(i/\hbar)\,\delta S\,(1-u^2)^{3/2} \cos(3\gamma)} \times
                                                                   \nonumber \\
      &   & \qquad \times \left\{ \frac{2|\Delta S|}{\hbar} A_C\, u^2 +
            \!\sqrt{\frac{4|\Delta S|}{\hbar\pi}}\,(1\!-\!u^2)
            \sqrt{\frac{2\pi|\delta S|}{\hbar}}
            \left[ {\bar A}_{AB} + \Delta A_{AB} \cos(3\gamma) \right]
            \right\}\!.
\end{eqnarray}
The second line above is again the parameterized product $A(u,\gamma)\,{\cal J}
(u,\gamma)$, with coefficients chosen such that asymptotically for large
$\Delta S$ and $\delta S$, the Gutzwiller limit
\begin{equation}\label{dghhgutz}
  \delta g_G = A_A \cos(S_A/\hbar-\sigma_A \pi/2)
             + A_B \cos(S_B/\hbar-\sigma_B \pi/2)
             + A_C \cos(S_C/\hbar-\sigma_C \pi/2)
\end{equation}
is reached. (The degeneracy factors due to discrete symmetries discussed above
are again included in the amplitudes.) In the limit $\Delta S=\delta S=0$, on
the other hand, $\M_u$ \eq{Mhhuni} still reduces to unity as it should.

The two-dimensional integrals in \eq{Mhhuni} cannot be done analytically here.
It turns out, however, that we can approximate them without violating the above
two limits.\footnote{These approximations correspond, in fact, to neglecting
only higher-order terms in $\hbar$. We have checked by numerical integration
that these do not affect the results discussed below.} For that we note that the
first term (corresponding to orbit C) asymptotically gets contributions only
from $u\simeq 1$; we may therefore put $u=1$ in the exponent of the $\gamma$
integral which then becomes unity. The second term of \eq{Mhhuni},
corresponding to the orbits A and B, asymptotically only gets contributions
from $u\simeq 0$. Putting $u=0$ in the exponents of the $\gamma$ integrals
leads to Bessel functions $J_0$ and $J_1$ like in the TGU formula \eq{trTGU}.
The remaining $u$ integrals can now be expressed again in terms of the Fresnel
functions after some partial integrations; hereby we keep only the
leading-order terms in $\hbar$. We then arrive at the uniform trace formula for
the HH potential, including the contributions from the three primitive orbits
A, B, and C,
\begin{eqnarray}\label{dghhuni}
\delta g_u & = & A_C \cos\!\left({S_C\over\hbar}+{\pi\over 2}\right) \nonumber\\
           & - & \sqrt{\left|\frac{2\delta S}{\Delta S}\right|} \left[
               {\bar A}_{AB}\,J_0\!\left(\frac{\delta S}{\hbar}\right)
               \cos\!\left({S_C\over\hbar}+{\pi\over 2}\right)
             - \Delta A_{AB}\,J_1\!\left(\frac{\delta S}{\hbar}\right)
               \sin\!\left({S_C\over\hbar}+{\pi\over 2}\right)\right]\nonumber\\
           & + & C\!\left(\!\sqrt{\frac{2|\Delta S|}{\hbar\pi}}\right)
                           \sqrt{\frac{4\pi|\delta S|}{\hbar}}\left[
                {\bar A}_{AB}\,J_0\!\left(\frac{\delta S}{\hbar}\right)
                \cos\!\left({{\bar S}_{AB}\over\hbar}\right)
             -  \Delta A_{AB}\,J_1\!\left(\frac{\delta S}{\hbar}\right)
                \sin\!\left({{\bar S}_{AB}\over\hbar}\right) \right]\nonumber\\
           & + & S\!\left(\!\sqrt{\frac{2|\Delta S|}{\hbar\pi}}\right)
                           \sqrt{\frac{4\pi|\delta S|}{\hbar}}\left[
                {\bar A}_{AB}\,J_0\!\left(\frac{\delta S}{\hbar}\right)
                \sin\!\left({{\bar S}_{AB}\over\hbar}\right)
             +  \Delta A_{AB}\,J_1\!\left(\frac{\delta S}{\hbar}\right)
                \cos\!\left({{\bar S}_{AB}\over\hbar}\right) \right]\!.
\end{eqnarray}
In the low-energy limit
$\Delta S\rightarrow 0$, $\delta S\rightarrow 0$, the second line of
\eq{dghhuni} cancels the orbit C term in the first line, the fourth line
vanishes, and the third line yields the HO trace formula \eq{ho2}.

At energies where the C orbit is well isolated and separated from the A and B
orbits ($\Delta S\gg \hbar$) but the splitting of A and B is still small
($\delta S \siml \hbar$), we can use the asymptotic forms \eq{CSasy} of the
Fresnel functions. Hereby the second terms from \eq{CSasy} combine to cancel
the second line in \eq{dghhuni}, whereas the leading terms combine into
\begin{eqnarray}\label{dghhullmo}
 \delta g_u & = & A_C \cos\!\left({S_C\over\hbar}+{\pi\over 2}\right)
                  + \sqrt{\frac{2\pi|\delta S|}{\hbar}}\times \nonumber \\
            &   & \quad \times \left[
                  {\bar A}_{AB}\,J_0\!\left(\frac{\delta S}{\hbar}\right)
                  \cos\!\left({{\bar S}_{AB}\over\hbar}-{\pi\over 4}\right)
              -   \Delta A_{AB}\,J_1\!\left(\frac{\delta S}{\hbar}\right)
                  \sin\!\left({{\bar S}_{AB}\over\hbar}-{\pi\over 4}\right)
                                                 \right]\!, \; (x\gg 1)
\end{eqnarray}
so that the second line contains nothing but the TGU approach to orbits
A and B, kept separate from the isolated orbit C contribution.

In \fig{hhudgps} we show the numerical results obtained with the uniform
approximation \eq{dghhuni} using $r_m=2$ (dashed line), compared with the
quantum-mechanical result (solid line) for the HH potential with $\alpha=0.04$
and a Gaussian averaging with width $\gamma=0.25 \,\hom$. The agreement is
perfect for all energies up to $e\sim 0.73$. The differences seen for $e\simg
0.73$ are due to some missing orbits which arise after period-doubling
bifurcations of the primitive orbits A and C, and perhaps to the isochronous
bifurcation of the A orbit that occurs only at $e\simeq 0.97$ but may make 
itself felt in the amplitude $A_A$ already at smaller energies. (Note that, in
contrast to the HH4 case above, we have not smoothed this amplitude here.) All
these bifurcations can be treated with the uniform approximations developed in
Refs.\ \cite{bif}. Part of the disagreement for $e\simg 0.75$ is also
due to inaccuracies in the diagonalisation of the quantum HH
Hamiltonian in a finite basis \cite{bblm}.

Like in the case of the HH4 potential, the SU(2) limit is reached in the
uniform approximation \eq{dghhuni} through a subtle cancellation of divergences
in the Gutzwiller amplitudes. Numerically, the most robust procedure is to use
the perturbative result \eq{Mhh} for values of $x=6\Delta S/7\hbar$ up to $\sim
1.5 - 2.5$, and then to switch to the uniform approximation.

\Figurebb{hhudgps}{150}{70}{795}{580}{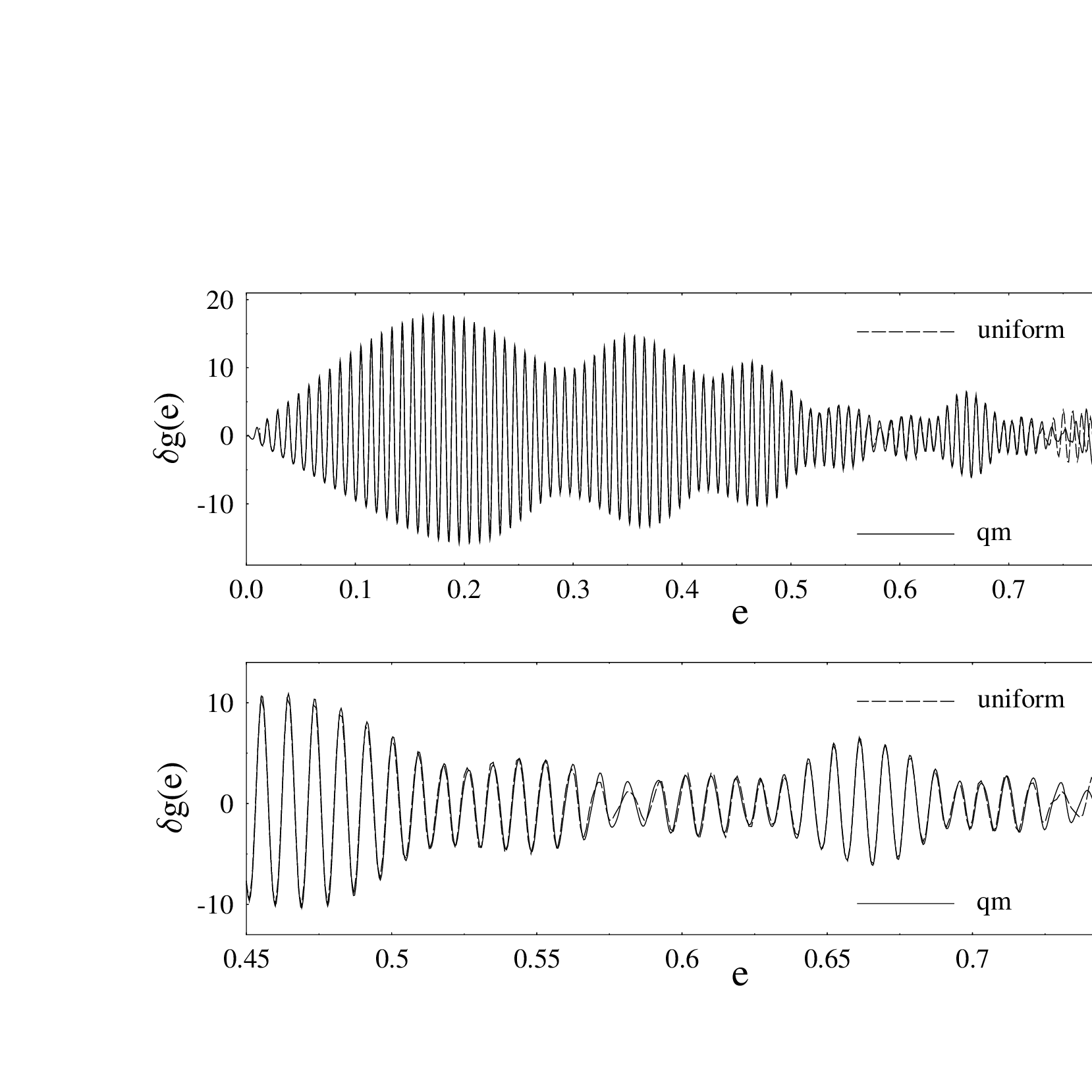}{11.0}{15.5}{
Oscillating part of level density of the H\'enon-Heiles potential
($\alpha=0.04$), Gaussian averaged over a range $\gamma=0.25$, versus
scaled energy $e$. Solid line: quantum-mechanical result; dashed line:
semiclassical result in the present uniform approximation with $r_m=2$. (Units:
$\hom$.)
}

\newpage

\section{Axial quadrupole deformations of a spherical cavity}

We finally discuss three-dimensional cavities with axially symmetric quadrupole
deformations that can be treated with the uniform approximation developed above
for the HH potential. The boundary of the cavities is in polar coordinates
given by
\begin{eqnarray}\label{caveps}
  R(\theta,\phi) = R_0\,[1+\epsilon P_2(\cos\theta)]\,,
\end{eqnarray}
where $P_2(x)=(3x^2-1)/2$ is the second Legendre polynomial. The periodic
orbits in a spherical cavity with ideally reflecting walls have been discussed
extensively by Balian and Bloch \cite{bb}, and an analytical trace formula has
been given by these authors for its level density which we again write as in
Eq.\ \eq{U1sym}. The three-dimensional degeneracy of the polygonal orbits with
$N\geq 3$ corners, due to the SO(3) symmetry of the sphere, can be described by
the three Euler angles ($\alpha,\beta,\gamma$), and the corresponding group
integral is \cite{gsym,crp}
\begin{equation}\label{so3int}
  \frac{1}{8\pi^2} \int_0^{2\pi} d\gamma \int_0^\pi \sin\beta \, d\beta
  \int_0^{2\pi} d\alpha = 1\, .
\end{equation}
We choose the Euler angles such that $\beta$ is the angle between the normal to
the plane of motion of an orbit family and the $z$ axis, $\gamma$ describes the
orientation of a single orbit within this plane (i.e., within the family), and
$\alpha$ describes rotations of the orbit plane around the $z$ axis. As long as
we restrict the deformations of the cavity to be axially symmetric (around the
$z$ axis), the action of the perturbed orbits will not depend on the angle
$\alpha$. For small deformations, the contribution of each periodic orbit
family to the spherical trace formula can thus again be written in the
perturbative approach as
\begin{equation} \label{trsph}
 \delta g_{pert} = \Re e \, \left\{ \M(x,y) e^{i(S_0/\hbar - \sigma_0 \pi/2)}
                                                                     \right\},
\end{equation}
where the modulation factor is
\begin{eqnarray}\label{Mso3ax}
 \M(x,y) & = & {1\over {2\pi}} \int_0^{2\pi} \!\! d\gamma \int_0^1 du\,
               e^{i\delta S(u,\gamma;x,y)}.
\end{eqnarray}
Here $x$ and $y$ are parameters depending on the deformation $\epsilon$,
the energy, and the specific orbit type. When both $x$ and $y$ are non-zero,
the orbit families are broken into singly degenerate families of orbits lying
in planes. These come in two types. One type is lying in planes perpendicular
to the symmetry axis (coming from the end-point correction at $u=1$); we will
henceforth call them the {\it equatorial orbits}. They are identical to the
orbits in a two-dimensional circular billiard.
The other type of orbits lie in planes that contain the symmetry axis
(coming from the stationary point $u=0$); we will call these the {\it planar
orbits}. They are isolated in their plane of motion and will, for small
deformations, come in pairs of stable and unstable orbits. With the
exception of a completely isolated linear orbit that oscillates along
the symmetry axis, all the
equatorial and planar orbits have the continuous one-parameter degeneracy
due to rotation of the plane of motion around the
symmetry axis.

We see therefore that the situation is completely analogous to that of the HH
potential discussed in the previous section: orbit C corresponds to the
equatorial orbit families (with two orientations connected to time reversal),
and the orbits A and B to the pairs of planar orbits. The only difference is
that, since we start here from a three-dimensional system with SO(3) symmetry,
the breaking of two degrees of degeneracy leaves the orbits in singly
degenerate families. Furthermore, the planar orbits do not have the discrete
three-fold rotational degeneracy as the A and B orbits in the HH potential, but
occasional two-fold degeneracies due to the symmetry of reflection at the
equator plane. We can therefore take over the previous results with little
effort. We shall first discuss the family of polygon orbits with $N\geq 3$
corners, and come to the diameter orbits ($N=2$) later.


\subsection{Polygon orbits with $N\geq 3$}

The families of polygon orbits in the sphere have in principle the full SO(3)
symmetry. However, when we limit the deformation of the cavity to
axially-symmetric shapes, the action shift depends only on two of the Euler
angles, so that the symmetry breaking problem is the same as in the HH
potential considered above. In first-order perturbation theory \cite{mbc},
the action shift depends only on the angle $\beta$:
\begin{equation}\label{polyshift}
\delta S_1 = - \Delta S \, P_2(\cos\beta)\,, \qquad
               \Delta S = \frac{\epsilon}{2} S_0\,.
\end{equation}
This corresponds to the fact that a quadrupole deformation to lowest order in
$\epsilon$ is identical to that of an axially symmetric ellipsoid, which is an
integrable system and thus has an extra symmetry with two-fold degenerate orbit
families (i.e., polygons with $N\ge 3$ fitting into the ellipse that gives the
boundary of the plane of motion). As for the orbits A and B of the HH system,
this symmetry will be broken only at a higher order in the perturbation
expansion; in the present case we expect this to happen at the third order in
$\epsilon$, with an action shift
\begin{equation}\label{polyshift3}
\delta S_3 = -\delta S_{pl}\, \sin^3\beta \cos(n\gamma)\,,
\end{equation}
where $n$ counts the discrete symmetries of the planar orbits and will again be
absorbed into the Gutzwiller amplitudes. With this, the modulation factor
becomes
\begin{equation}\label{Mquad}
   \M(x,y)  =  \int_0^1 du \, e^{ix(1-3u^2)/2}\,
               {1\over {2\pi}} \int_0^{2\pi} \!\! d\gamma \,
               e^{-iy(1-u^2)^{3/2} \cos(n\gamma)}
\end{equation}
with $\hbar x=\Delta S$ and $\hbar y=\delta S_{pl}$. For small $x$ and $y$, the
action shifts will be
\begin{eqnarray}\label{quashift1}
    (u=1): & \quad & S_{eq} - S_0 = -x = -\hbar \Delta S\,, \nonumber \\
    (u=0): & \quad & {\bar S}_{pl} - S_0 = +x/2 = +\hbar \Delta S/2\,,
\end{eqnarray}
with
\begin{eqnarray}\label{quashift2}
    {\bar S}_{pl} = \frac12 (S_{pl}^u + S_{pl}^s) \,, \quad
         S_{pl}^u = {\bar S}_{pl} + \delta S_{pl} \,, \quad
         S_{pl}^s = {\bar S}_{pl} - \delta S_{pl} \,.
\end{eqnarray}
The modulation factor \eq{Mquad} has exactly the same form as that of the HH
case \eq{Mhhint2}. However, different from there, $x$ and $y$ can have both
signs, depending on the sign of $\epsilon$. Therefore, switching from prolate
deformations ($\epsilon>0$), which is analogous to the HH situation, to oblate
deformations ($\epsilon<0$), one has to take the complex conjugate of the
modulation factor before using it in \eq{trsph}. This only affects some of the
signs in the final results; in the following formulae the upper signs
correspond to the prolate case and the lower signs to the oblate case.

The final trace formula for one type of orbits (with fixed number $N\geq 3$ of
corners, and without including their higher repetitions) becomes
\begin{eqnarray}\label{dgqu}
\delta g_u^{(N\geq3)} & = & A_{eq} \cos\Phi_{eq}
                 - \sqrt{\frac{2|\delta S|}{|\Delta S|}} \left[
                   {\bar A}_{pl}\,J_0\!\left(\frac{\delta S}{\hbar}\right)
                                 \cos\Phi_{eq}
                 - \Delta A_{pl}\,J_1\!\left(\frac{\delta S}{\hbar}\right)
                                 \sin\Phi_{eq} \right]     \nonumber \\
           & + & C\!\left(\!\sqrt{\frac{2|\Delta S|}{\pi\hbar}}\right)
                   \sqrt{\frac{4\pi|\delta S|}{\hbar}} \left[
                 {\bar A}_{pl}\,J_0\!\left(\frac{\delta S}{\hbar}\right)
                                 \cos{\bar \Phi}_{pl}
               - \Delta A_{pl}\,J_1\!\left(\frac{\delta S}{\hbar}\right)
                                 \sin{\bar \Phi}_{pl} \right] \nonumber \\
         & \pm & S\!\left(\!\sqrt{\frac{2|\Delta S|}{\pi\hbar}}\right)
                   \sqrt{\frac{4\pi|\delta S|}{\hbar}}\, \left[
                 {\bar A}_{pl}\,J_0\!\left(\frac{\delta S}{\hbar}\right)
                                 \sin{\bar \Phi}_{pl}
               + \Delta A_{pl}\,J_1\!\left(\frac{\delta S}{\hbar}\right)
                                 \cos{\bar \Phi}_{pl}         \right],
\end{eqnarray}
where we have redefined
\begin{eqnarray}
 \frac32 \hbar x = \Delta S = {\bar S}_{pl} - S_{eq}\,,       & \quad &
 \hbar y = \delta S = \frac12(S_{pl}^u - S_{pl}^s)\,,           \nonumber \\
 \Phi_{eq} = {S_{eq}\over\hbar} - \sigma_{eq}{\pi\over 2},    & \quad &
 {\bar\Phi}_{pl} = {{\bar S}_{pl}\over\hbar} - \sigma_0 \frac{\pi}{2}\,.
\end{eqnarray}
and the combinations of Gutzwiller amplitudes (including discrete
degeneracy factors)
\begin{equation}
   {\bar A}_{pl} = \frac12 (A_{pl}^u + A_{pl}^s)\,, \quad
   \Delta A_{pl} = \frac12 (A_{pl}^u - A_{pl}^s)\,.
\end{equation}
The correct Maslov indices of the asymptotic orbits become
\begin{eqnarray}\label{maslov}
 \sigma_{eq}=\sigma_0 \mp 1\,, \quad
 {\bar\sigma}_{pl}=\frac12\,(\sigma_{pl}^u+\sigma_{pl}^s) = \sigma_0\pm\frac12,
 \quad \sigma_{pl}^u=\sigma_{pl}^s+1.
\end{eqnarray}

At moderately large deformations, the equatorial and planar orbits are
sufficiently well separated, so that we can take the limit $|\Delta S|\gg\hbar$.
The result \eq{dgqu} then simplifies to
\begin{eqnarray}\label{dgqul}
\delta g_u^{(N\geq3)} & = & A_{eq} \cos\Phi_{eq}               \nonumber \\
           & + & \sqrt{\frac{2\pi|\delta S|}{\hbar}}    \left[
                 {\bar A}_{pl}\,J_0\!\left(\!\frac{\delta S}{\hbar}\!\right)
                 \cos\!\left(\frac{{\bar S}_{pl}}{\hbar}-
                                     {\bar\sigma}_{pl}\frac{\pi}{2}\right)
               - \Delta A_{pl}\,J_1\!\left(\!\frac{\delta S}{\hbar}\!\right)
                 \sin\!\left(\frac{{\bar S}_{pl}}{\hbar}-
                                     {\bar\sigma}_{pl}\frac{\pi}{2}\right)
                                                       \right],
\end{eqnarray}
which again contains the simple TGU formula applied to the planar orbits. At
larger deformations (but still before any bifurcations take place), we expand
the above for $|\delta S|\gg\hbar$ and obtain the Gutzwiller limit
\begin{equation}
\delta g_G^{(N\geq3)} = A_{eq} \cos\Phi_{eq} + A^u_{pl} \cos\Phi^u_{pl}
           + A^s_{pl} \cos\Phi^s_{pl}.
\end{equation}

\subsection{Diameter orbits $(N=2)$}

The diameter orbits ($N=2$) in a spherical cavity have only a continuous
degeneracy of two, since rotation about themselves does not bring about any new
orbit \cite{bb}. It is convenient to redefine the Euler angles such that $\beta$
describes the angle between the diameter orbit and the $z$ axis. In first-order
perturbation theory, the action shift of the diameter orbit due to a quadrupole
deformation \eq{caveps} is then given by \cite{mbc}
\begin{eqnarray}\label{diashift}
\delta S_1 = \epsilon S_0 P_2(\cos\beta)
\end{eqnarray}
Hence the modulation factor for small deformations becomes
\begin{equation}\label{Mqdia}
   \M(x)  =  \int_0^1 du \, e^{ix(3u^2-1)/2}
\end{equation}
with $\hbar x = \epsilon S_0$. Asymptotically, we get from $u=0$ the equator
orbits which keep the U(1) degeneracy corresponding to rotation about the
$z$ axis, and from the end point $u=1$ we find the isolated diameter orbit along
the $z$ axis. Their action shifts for small $x$ will be
\begin{eqnarray}\label{actdia}
S_{eq}-S_0 = - \hbar x/2\,, \quad S_{iso}-S_0 = +\hbar x
\end{eqnarray}
The situation thus corresponds exactly to the case for the polygonal orbits,
but with $y=0$ since there is no further splitting of the isolated diameter
orbit, and with the roles of prolate and oblate deformations interchanged. The
uniform trace formula for the diameter contributions is thus
\begin{eqnarray}\label{dgudia}
\delta g_u^{(N=2)} & = & \left( A_{iso} - \sqrt{\frac{\hbar}{\pi|\Delta S|}}\,
                         A_{eq} \right) \cos\Phi_{iso}            \nonumber \\
           & + & \sqrt{2}\,A_{eq}\left[
               C\!\left(\!\sqrt{\frac{2|\Delta S|}{\pi\hbar}}\right)
                \cos\!\left(\frac{S_{eq}}{\hbar}-\sigma_0\frac{\pi}{2}\right)
           \mp S\!\left(\!\sqrt{\frac{2|\Delta S|}{\pi\hbar}}\right)
                \sin\!\left(\frac{S_{eq}}{\hbar}-\sigma_0\frac{\pi}{2}\right)
                 \right].
\end{eqnarray}
Hereby we have (re)defined
\begin{eqnarray}\label{diadiv}
\Delta S = S_{iso}-S_{eq}, \quad
\Phi_{iso} = \frac{S_{iso}}{\hbar}-\sigma_{iso}\,\frac{\pi}{2}, \quad
\sigma_{iso} & = \sigma_0\pm 1\,.
\end{eqnarray}
For $|\Delta S|\gg\hbar$, we get the Gutzwiller limit for the diameter orbits:
\begin{equation}
\delta g_G^{(N=2)} = A_{eq} \cos\Phi_{eq} + A_{iso} \cos\Phi_{iso}
\end{equation}
with
\begin{eqnarray}\label{diadivg}
\Phi_{eq} = \frac{S_{eq}}{\hbar}-\sigma_{eq}\,\frac{\pi}{2}, \qquad
\sigma_{eq} = \sigma_0\mp \frac12.
\end{eqnarray}
In all equations above, the upper and lower signs correspond to $\Delta S > 0$
(prolate case) and $\Delta S < 0$ (oblate case), respectively.

In \fig{qudg01} we show the oscillating part $\delta g(k)$ of the level density
for a very small quadrupole deformation of $\epsilon=0.01$, Gaussian-averaged
over $k$ with a range $\gamma=0.6\,R^{-1}$ and plotted versus the wave number
$k=\sqrt{2mE}/\hbar$. Note the pronounced ``supershell'' oscillations that are
mainly a result of the interfering triangle and square orbits \cite{bb}. In the
semiclassical result (dashed line), periodic orbits with up to $N=6$
reflections were included in the uniform approximation given by Eqs.\
\eqq{dgqu}{dgudia}; identical results are obtained at this deformation in the
perturbative approach. Both reproduce very well the quantum-mechanical result
(solid line). For details concerning the calculation of the Gutzwiller
amplitudes for the equatorial and planar orbits in non-integrable axially
deformed cavities, we refer to a forthcoming publication \cite{mesb}. 

\Figurebb{qudg01}{75}{35}{567}{275}{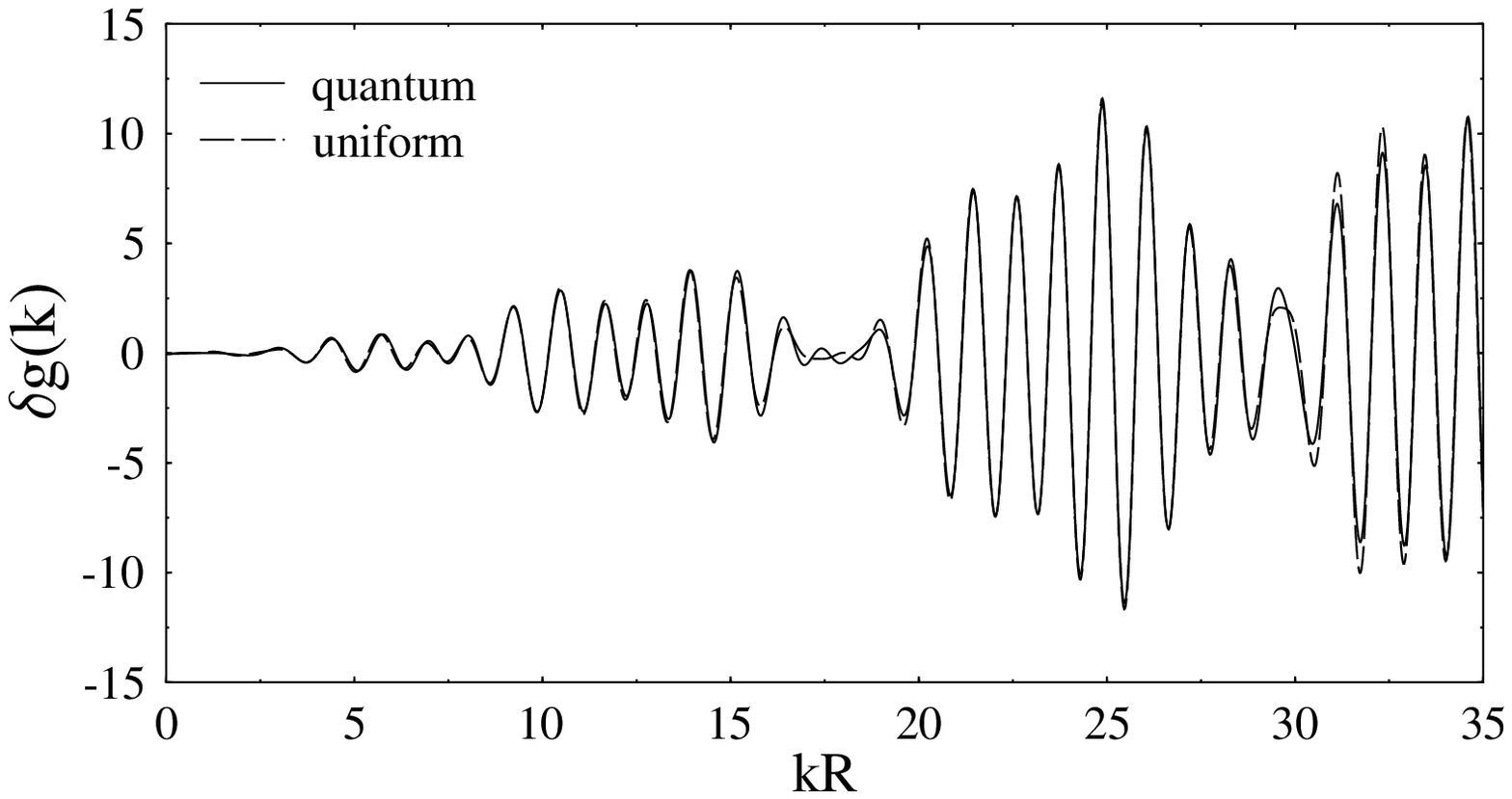}{7.2}{15.0}{
Oscillating part of level density in an axially-symmetric quadrupole-deformed
cavity with deformation $\epsilon=0.01$, Gaussian averaged over $k$ with a
range $\gamma=0.6$, versus wave length $k$ (units: $R^{-1}$). Solid line:
quantum-mechanical result; dashed line: semiclassical result in the present
uniform approximation using equatorial and planar orbits with up to $N=6$
reflections.
}

In \fig{qudg1}, we show the corresponding results at a quadrupole deformation of
$\epsilon=0.1$ at which the supershell beating is already reduced \cite{mbc}.

\Figurebb{qudg1}{75}{35}{567}{275}{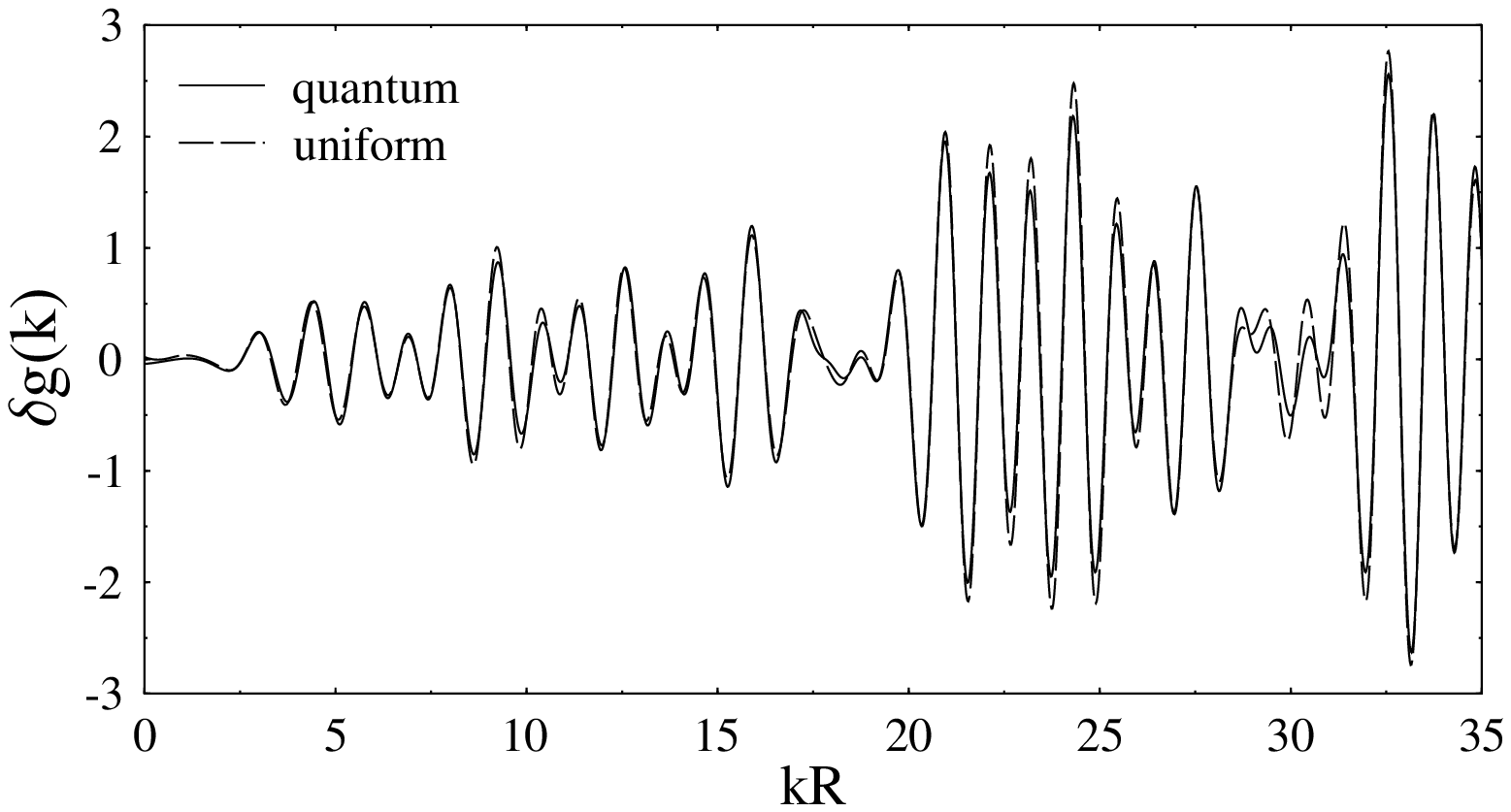}{7.2}{15.0}{
Same as \fig{qudg01}, but for the deformation $\epsilon=0.1$.
}

The relative importance of the different orbits at these deformations can most
easily be analyzed in the Fourier spectra of the oscillating level density
$\delta g(k)$. Since billiard systems exhibit scaling (i.e., the properties of
the periodic orbits do not depend on energy), the Fourier 

\Figurebb{quft01}{70}{460}{576}{690}{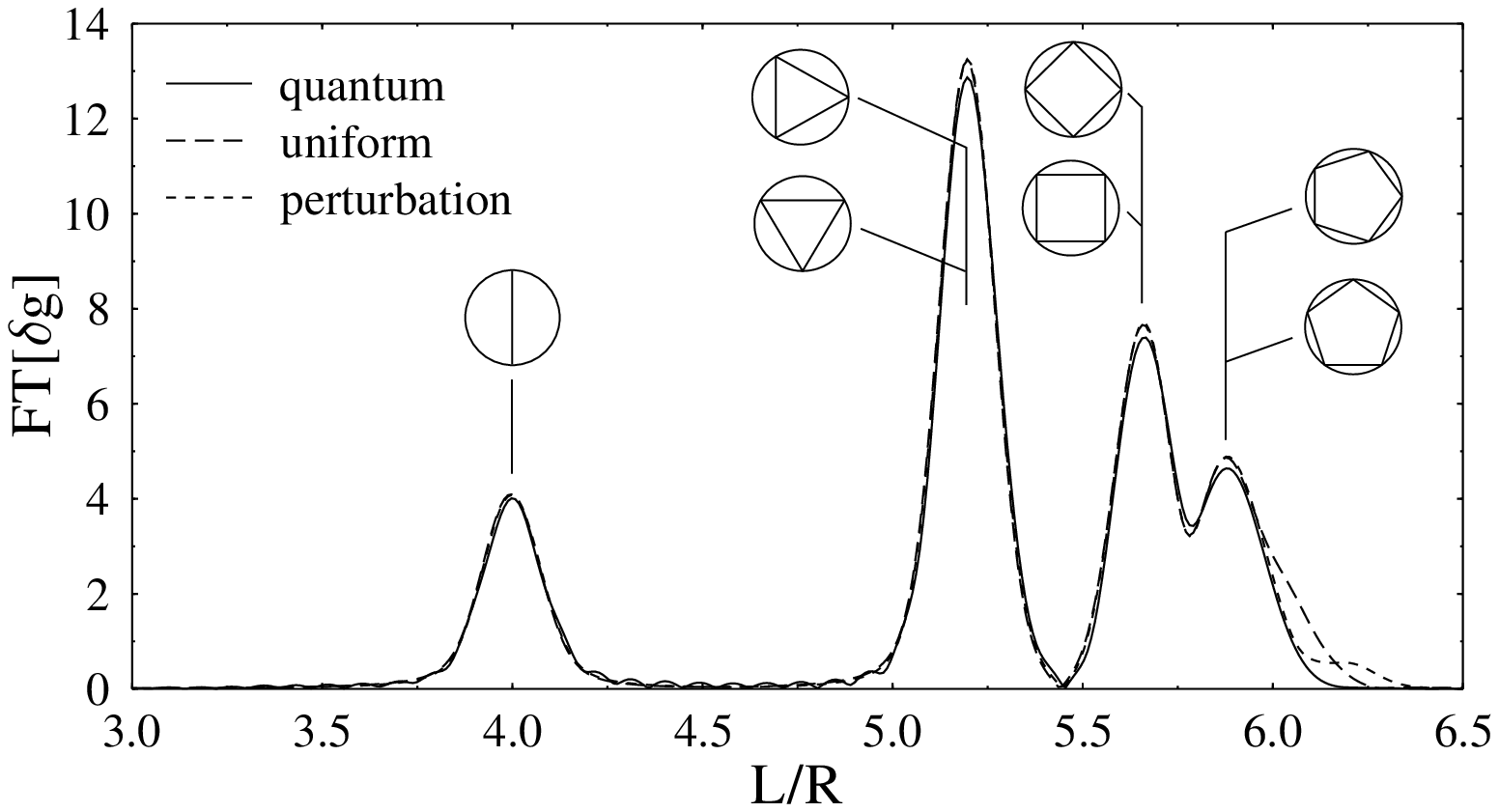}{7.1}{15.0}{
Fourier transform of the level densities shown in \fig{qudg01}. The
symbols indicate the periodic orbits corresponding to the Fourier peaks. The
deformation ($\epsilon=0.01$) is so small here that equatorial and planar
orbits are not separated yet. The perturbation approach (short-dashed line)
gives practically identical results as the uniform approximation (long-dashed
line) and the quantum mechanics (solid line).
}

\Figurebb{quft1}{70}{537}{526}{747}{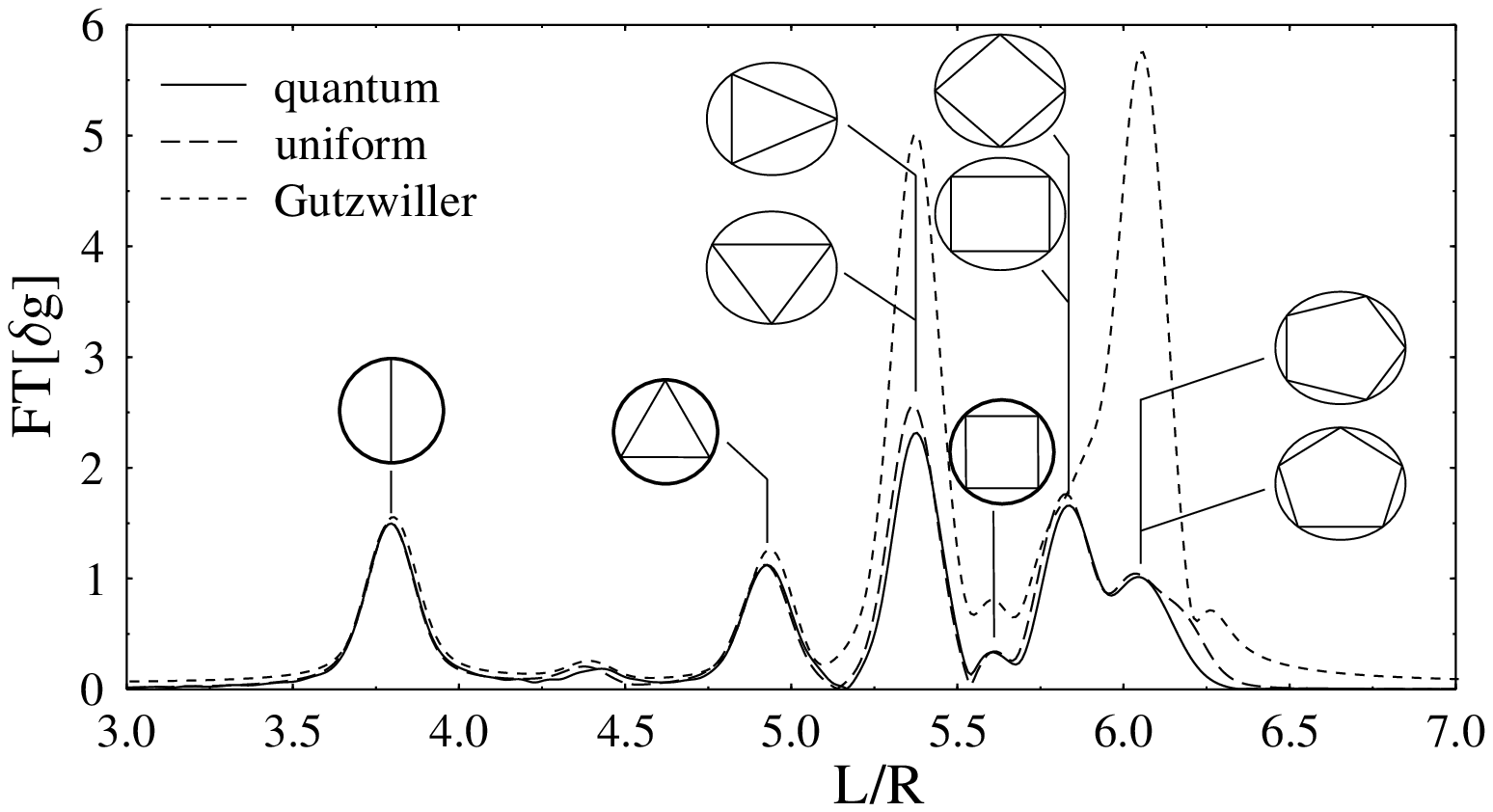}{6.5}{15.0}{
Fourier transform of the level densities shown in \fig{qudg1}. At this
deformation ($\epsilon=0.1$), the equatorial orbits (symbols in heavy circles)
are well separated from the planar orbits (symbols in thin ovals). However, the
pairs of isolated stable and unstable planar orbits are not separated yet, as
seen from the large peaks obtained with the Gutzwiller trace formula
(thin-dashed line). The tiny bump at $L \simeq 4.4\,R$ corresponds to the
unstable isolated diameter orbit along the symmetry axis, whose Gutzwiller
amplitude is smaller than all the others by a factor $\sqrt{\hbar/kR}$.
}

\noindent
transform with
respect to $k$ gives directly power spectra in which the peaks occur at the
lengths of the contributing periodic orbits. Figures \ref{quft01} and
\ref{quft1} present the absolute values of the Fourier transforms of the
level densities shown in the above two figures. In \fig{quft01}, the peaks
corresponding to equatorial and planar orbits cannot be separated, since at the
small deformation $\epsilon=0.01$ the spherical tori are hardly broken. Indeed,
the perturbative result (short-dashed line) gives here practically identical
results as the uniform approximation (long-dashed line); both agree very well
with the quantum result. At the deformation $\epsilon=0.1$, the equatorial and
planar orbits are already well separated, as can be seen from \fig{quft1}, so
that the full uniform approximation \eq{dgqu} for the orbits with $N\ge 3$ can
here be replaced by the TGU approximation given in Eq.\ \eq{dgqul}. However,
the separation of the stable and unstable planar orbits is still very small.
Therefore, the Gutzwiller trace formula dramatically overestimates their
combined amplitudes, as seen from the short-dashed line.


\section{Summary}

We have derived uniform approximations for the semiclassical description of
systems with perturbed SU(2) and SO(3) symmetry. Different from the case of
bifurcations, where uniform approximations are most effectively derived from
the expansion of the classical action in phase-space into normal forms
\cite{bif}, we use the group integral representation of the unperturbed trace
formula \cite{gsym} and the classical perturbation theory \cite{crp} as
starting points for our development. 

In terms of analytical trace formulae we can interpolate smoothly from the
integrable limits to the limits where their symmetries are broken and the 
asymptotic Gutzwiller trace formulae for the leading periodic orbits are 
reached. In the two-dimensional H\'enon-Heiles type potentials with broken
SU(2) symmetry, these orbits are isolated, whereas in the axially-symmetric
quadrupole-deformed cavities with broken SO(3) symmetry, they retain a
one-dimensional degeneracy corresponding to the rotation about the symmetry
axis. In all cases, the few shortest periodic orbits that were considered could
account quantitatively for the gross-shell effects found in the coarse-grained
quantum-mechanical level densities.

Unlike for the breaking of an orbit family with U(1) symmetry into pairs 
of stable and unstable isolated orbits, for which case Tomsovic {\it et al.}\ 
\cite{TGU} have found the universal uniform approximation given in Eq.\ 
\eq{trTGU}, our present results for SU(2) and SO(3) breaking are not universal. 
This is due to more complicated scenarios for the breaking of 
rational tori that arise when more degrees of freedom, or
higher-dimensional degeneracies of the orbit families, are involved. The two 
examples of the quartic and cubic H\'enon-Heiles potentials show how 
the symmetry breaking 
can happen at different orders in the perturbation theory and lead to quite 
different modulation factors; see Eqs.\ \eq{M441} and
\eq{Mhhint2} which were the starting point of our development.
However, due to the feasibility of the perturbation theory and the
generality of the scheme for the uniform approximation used in this study, 
our method can easily be applied also to other potentials and different 
symmetries of the unperturbed system.  

\bs

We acknowledge a critical reading of the manuscript by Martin Sieber, and
further helpful discussions with Ch.\ Amann, J. Blaschke, S. Creagh, S.
Fedotkin, J. Law, and S. Tomsovic. This work has been partially
supported by Deutsche Forschungsgemeinschaft (grant No.\ Br~733/9-1) 
and by Deutscher Akademischer Austauschdienst (DAAD).

\begin{appendix}

\section{Analytical results for linear periodic orbits}

For the test of numerical routines that solve the equations of motion and
determine the actions of periodic orbits, it might be helpful to compare with
analytical results where available. The straight-line librating periodic orbits
in the H\'enon-Heiles potentials allow for an analytic calculation of their
actions or periods. We give here the results and a brief sketch of their
derivations.

\subsection{The A$_1$ and A$_2$ orbits in the quartic H\'enon-Heiles potential}

The potential
\begin{equation}
  V(r,\theta) = \frac{1}{2}\,\omega^2 r^2-{\alpha\over 4}\, r^4\cos(4\theta)\,,
\end{equation}
has two pairs of linear librating orbits: the A$_1$ orbits oscillating along
the $x$ and $y$ axes ($\theta=0$ and $\pi/2$), and the orbits A$_2$ oscillating
along the diagonals ($\theta=\pi/4$ and $3\pi/4$). We scale the potential with
the factor $1/E^*=4\alpha/\omega^4$, so that the equations for their classical
turning points are
\be \label{V4tp}
  e = 2x^2 \mp x^4
\ee
in terms of the scaled energy $e$ \eq{H4scal} and the scaled radial coordinate
$x=r\sqrt{\alpha}/\omega$. The upper and lower sign in \eq{V4tp} holds for the
orbits A$_1$ and A$_2$, respectively. The four solutions of Eq.\ \eq{V4tp} with
the ``$-$'' sign are given by $\pm x_1$ and $\pm x_2$, where
\be
 x_1=\sqrt{1-\sqrt{1-e}}, \quad x_2=\sqrt{1+\sqrt{1-e}},
\ee
and $\pm x_1$ are the classical turning points of the A$_1$ orbits. Their
action is then
\be
S_{A_1} = 8\sqrt{2}\,\,\frac{E^*}{\omega}\int_0^{x_1}\!\!\sqrt{(e-2x^2+x^4)}\,dx
        = 8\sqrt{2}\,\,\frac{E^*}{\omega}\int_0^{x_1}\!\!
                                              \sqrt{(x_1^2-x^2)(x_2^2-x^2)}\,dx
\ee
which can be expressed in term of the complete elliptic integrals ${\bf
K}(t)=F(\pi/2,t)$ and ${\bf E}(t)=E(\pi/2,t)$ in terms of the quantity
\be
  t = \left(\frac{x_1}{x_2}\right)^2 = \frac{1-\sqrt{1-e}}{1+\sqrt{1-e}}.
\ee
The result is
\be \label{SA1}
 S_{A_1}(e) = \frac{16\sqrt{2}}{3}\, \frac{E^*}{\omega} \,\sqrt{1+\sqrt{1-e}}\,
              \left[{\bf E}(t)-\sqrt{1-e}\,{\bf K}(t)\right]. \qquad (e<1)
\ee
At $e=1$ we get simply $S_{A_1}(1)=16\sqrt{2}\,E^*/(3\omega)$.

For the A$_2$ orbits, the solutions of Eq.\ \eq{V4tp} with the ``$+$'' sign are
given by $\pm x_1$ and $\pm i x_2$, where now,
\be
 x_1=\sqrt{\sqrt{1+e}-1}, \quad x_2=\sqrt{\sqrt{1+e}+1},
\ee
and $\pm x_1$ are again the classical turning points. The action of the A$_2$
orbits is then
\be
   S_{A_2} = 8 \sqrt{2}\,\,\frac{E^*}{\omega}
               \int_0^{x_1} \!\!\sqrt{(x_2^2+x^2)(x_1^2-x^2)}\,dx
\ee
which becomes
\be \label{SA2}
  S_{A_2}(e) = \frac{16}{3}\,\frac{E^*}{\omega}\, (1+e)^{1/4}\,
               \left[\left(1+\sqrt{1+e}\right){\bf K}(\kappa)
                     -2\,{\bf E}(\kappa)\right],
\ee
in terms of the quantity
\be
  \kappa = \frac{x_1^2}{(x_1^2+x_2^2)}
         = \frac{\sqrt{1+e}-1}{2\sqrt{1+e}}.
\ee
Taylor expansion of Eqs.\ \eqq{SA1}{SA2} in powers of $e$ leads to the
result given in Eq.\ \eq{SAexp}.


\subsection{The A orbit in the standard H\'enon-Heiles potential}

The potential
\begin{equation}
  V(r,\theta) = \frac{1}{2}\,\omega^2 r^2-{\alpha\over 3}\, r^3\cos(3\theta)
\end{equation}
has linear librating orbits A oscillating along the symmetry axes ($\theta=0$,
$2\pi/3$ and $4\pi/3$). We scale the potential with the factor $1/E^* =
6\alpha^2/\omega^6$, so that the equation for the classical turning points is
\be \label{V3tp}
  e = 3x^2 - 2x^3
\ee
in terms of the scaled energy $e=E/E^*$ and the scaled radial coordinate
$x=r\alpha/\omega^2$. The real solutions for this cubic equation for $e\le 1$
are, with $x_1 \le x_2 \le x_3$,
\be
x_1 = \frac12 - \cos(\pi/3-\phi/3)\,,\qquad
x_2 = \frac12 - \cos(\pi/3+\phi/3)\,,\qquad
x_3 = \frac12 + \cos(\phi/3)\,,
\ee
where
\be
\cos\phi = 1 - 2e\,.
\ee
The action of the A orbit is then
\be
   S_A = 4\sqrt{3}\,\,\frac{E^*}{\omega}
         \int_{x_1}^{x_2} \!\! \sqrt{e-3x^2+2x^3}\,dx
       = 4\sqrt{6}\,\,\frac{E^*}{\omega}
         \int_{x_1}^{x_2} \!\! \sqrt{(x-x_1)(x_2-x)(x_3-x)}\,dx\,,
\ee
but we could not find an analytical expression for this integral. Instead, we
calculate the period $T_A=dS_A/dE$
\be
     T_A = \frac{\sqrt{6}}{\omega}
           \int_{x_1}^{x_2} \! \frac{1}{\sqrt{(x-x_1)(x_2-x)(x_3-x)}}\,dx
\ee
which again can be expressed in terms of a complete elliptic integral by
\be  \label{TA}
     T_A(e) = \frac{\sqrt{6}}{\omega}\,\frac{2}{\sqrt{x_3-x_1}}\,{\bf K}(q)\,,
              \qquad \qquad (e<1)
\ee
where
\be
     q = \left(\frac{x_2-x_1}{x_3-x_1}\right).
\ee
(Note that $T_A$ diverges at $e=1$.) Expansion of \eq{TA} in powers of $e$ and 
integrating over the energy $E$ leads to the result given in Eq.\ \eq{SAHexp}.

\end{appendix}


\begin{thebibliography}{10}

\setlength{\itemsep}{-0.25ex}

\bibitem{gutz} M. C. Gutzwiller, J. Math. Phys. {\bf 12}, 343 (1971);
               {\it Chaos in Classical and Quantum Mechanics}
               (Springer Verlag, New York, 1990).

\bibitem{bb}   R. Balian, and C. Bloch, Ann. Phys. (N. Y.) {\bf 69}, 76 (1972).

\bibitem{bt}   M. V. Berry and M. Tabor, Proc. R. Soc. Lond. {\bf A 349},
               101 (1976); {\it ibid.} {\bf A 356}, 375 (1977).

\bibitem{book} M. Brack and R. K. Bhaduri: {\it Semiclassical Physics},
               Frontiers in Physics Vol.\ {\bf 96} (Addison-Wesley,
               Reading, 1997).

\bibitem{stma} V. M. Strutinsky and A. G. Magner, Sov. J. Part. Nucl. {\bf 7},
               138 (1976).

\bibitem{gsym} S. C. Creagh and R. G. Littlejohn, Phys. Rev. A {\bf 44}, 836
               (1991); J. Phys. A {\bf 25}, 1643 (1992).

\bibitem{rich} P. J. Richens, J. Phys. {\bf A 15}, 2101 (1982).

\bibitem{bif}  M. Sieber, J. Phys. A {\bf 29}, 4716 (1996); M. Sieber,
               J. Phys. {\bf A 30}, 4563 (1997); H. Schomerus and M. Sieber,
               {\bf A 30}, 4537 (1997).

\bibitem{siel} M. Sieber, J. Phys. {\bf A 30}, 4563 (1997).

\bibitem{ozha} A. M. Ozorio de Almeida and J. H. Hannay, J. Phys. {\bf A 20},
               5873 (1987); see also A. M. Ozorio de Almeida: {\it Hamiltonian
               Systems: Chaos and Quantization} (Cambridge University Press,
               Cambridge, 1988).

\bibitem{mon}  A. G. Magner, S. N. Fedotkin, F. A. Ivanyuk, P. Meier, M. Brack,
               S. M. Reimann, and H. Koizumi, Ann. Physik (Leipzig) {\bf 6},
               555 (1997); A. G. Magner, S. N. Fedotkin, F. A. Ivanyuk,
               P. Meier, and M. Brack, Czech. J. Phys. {\bf 48}, 845 (1998).

\bibitem{OdA}  Ozorio de Almeida, in {\it Quantum Chaos and Statistical
               Nuclear Physics} (T. H. Seligmann and H. Nishioka, Eds.)
               Lecture Notes in Physics {\bf 263}, p.\ 197 (Springer Verlag,
               New York/Berlin, 1986).

\bibitem{URJ}  D. Ullmo, K. Richter and R. A. Jalabert, Phys. Rev. Lett.
               {\bf 74}, 383 (1995); see also K. Richter, D. Ullmo and R. A.
               Jalabert, Phys. Rep. {\bf 276}, 1 (1996).

\bibitem{crp}  S. C. Creagh, Ann. Phys. (N. Y.) {\bf 248}, 60 (1996).

\bibitem{mbc}  P. Meier, M. Brack and S. C. Creagh, Z. Phys. {\bf D 41},
               281 (1997).

\bibitem{clus} M. Brack, J. Blaschke, S. C. Creagh, A. G. Magner, P. Meier,
               and S. M. Reimann, Z. Phys. {\bf D 40}, 276 (1997).

\bibitem{kaor} K. Tanaka, S. C. Creagh and M. Brack, Phys. Rev. B {\bf 53},
               16050 (1996).

\bibitem{bcl}  M. Brack, S. C. Creagh, and J. Law, Phys. Rev. {\bf A 57},
               788 (1998).

\bibitem{TGU}  S. Tomsovic, M. Grinberg and D. Ullmo, Phys. Rev. Lett
               {\bf 75}, 4346 (1995); 
               see also D. Ullmo, M. Grinberg and S. Tomsovic,
               Phys. Rev. {\bf E 54}, 135 (1996).

\bibitem{nuc}  V. M. Strutinsky, A. G. Magner, S. R. Ofengenden,
               and T. D{\o}ssing, Z. Phys. {\bf A 283}, 269 (1977).

\bibitem{fis}  M. Brack, S. M. Reimann, and M. Sieber, Phys. Rev. Lett.
               {\bf 79}, 1817 (1997); see also M. Brack, P. Meier,
               S. M. Reimann, and M. Sieber, in: {\it Similarities and
               differences between atomic nuclei and clusters}, eds.\ 
               Y. Abe {\it et al.} (American Institute of Physics,
               1998) p.\ 17.

\bibitem{nis}  H. Nishioka, K. Hansen and B. R. Mottelson, Phys. Rev. {\bf B
               42}, 9377 (1990).

\bibitem{kao2} K. Tanaka, Ann. Phys. (N.Y.) {\bf 268}, 31 (1998).

\bibitem{qdot} S. M. Reimann, M. Persson, P. E. Lindelof, and M. Brack,
               Z. Phys. {\bf B 101}, 377 (1996).

\bibitem{mesb} P. Meier, M. Sieber, and M. Brack, to be published.

\bibitem{bj}   M. Brack and S. R. Jain, Phys. Rev. {\bf A 51}, 3462 (1995).

\bibitem{hh}   M. H\'{e}non and C. Heiles, Astr. J. {\bf 69}, 73 (1964).

\bibitem{bblm} M. Brack, R. K. Bhaduri, J. Law, M. V. N. Murthy, and Ch.
               Maier, Chaos {\bf 5}, 317, 707(E) (1995).

\bibitem{abst} M. Abramowitz and I. Stegun: {\it Handbook of mathematical
               functions} (Dover, N.Y., 1965).

\end{thebibliography}
\end{document}